\documentclass[aps,twocolumn,groupedaddress,reprint,amsmath,amssymb]{revtex4-2}
\usepackage[utf8]{inputenc}
\usepackage[T1]{fontenc}
\usepackage{amsmath}
\usepackage{amsfonts}
\usepackage{amssymb}
\usepackage{graphics,graphicx}
\usepackage{graphicx}
\usepackage{caption}
\usepackage{subcaption} 
\usepackage{wrapfig} 
\usepackage{epstopdf}
\usepackage{xcolor}
\usepackage{float}
\graphicspath{{figuras/}}

\usepackage{ulem}
\usepackage[breaklinks=true]{hyperref}
\usepackage{setspace}
\usepackage{graphicx}
\usepackage{color}
\usepackage{placeins}

\definecolor{darkraspberry}{rgb}{0.53,0.15,0.34}

\begin{document}

\title{Multicritical points of gravitational solitons and a black hole in four dimensions.}
\author{Mois\'es Bravo-Gaete}
\email{moisesbravog@gmail.com, mbravo@ucm.cl}
\affiliation{Departamento de Matem\'atica, F\'isica y Estad\'istica, Facultad de Ciencias
B\'asicas, Universidad Cat\'olica del Maule, Casilla 617, Talca, Chile.}

\author{Moaathe Belhaj Ahmed} \email{hoothyy789@gmail.com}
 \author{Robert B. Mann}
\email{rbmann@uwaterloo.ca}
\affiliation{Department of Physics and Astronomy, University of Waterloo, Waterloo, Ontario, N2L 3G1, Canada.}

\author{Constanza Quijada}
\email{constanza.quijada@unab.cl}
\affiliation{Departamento de F\'isica y Astronom\'ia, Universidad Andr\'es Bello, Autopista Concepcion-Talcahuano 7100, Talcahuano, Chile.}

\date{\today}          
\begin{abstract}
We present the first realization of multicritical points in four-dimensional general relativity, specifically within the context of Pleba\'nski nonlinear electrodynamics, using a polynomial structural function denoted as $\mathcal{H}(P)$. We show that this construction provides a systematic mechanism to engineer multicritical behavior in gravitational systems. By establishing an explicit mapping between matter theories expressed as power series in the Maxwell invariant 
$F$ and the Pleba\'nski formulation, we construct new families of electrically charged asymptotically anti-de Sitter black holes and magnetically charged solitons. In the grand-canonical ensemble, we analyze their thermodynamic properties and uncover a rich phase structure. We demonstrate that the soliton sector develops multiple swallowtail structures, signaling first-order phase transitions and allowing the coexistence of several magnetically charged solitons with a single electrically charged black hole. These configurations define multicritical points that generalize previously known triple points. We further show that the number of coexisting phases is controlled by the degree of the polynomial structural function, providing a direct link between the nonlinear electrodynamics couplings and the thermodynamic phase structure. In contrast, the black hole branch does not display swallowtail behavior, and it does not allow multiple electrically charged black holes to coexist with a magnetically charged soliton. 
\end{abstract}

\maketitle
\section{Introduction}\label{intro}
In thermodynamic systems, variations of external control parameters induce qualitative changes in the equilibrium configuration, driving the system into a more stable state. This phenomenon defines what is generally referred to as a phase transition (PT) \cite{Natsuume:2014sfa}. 
Asymptotically  anti-de Sitter (AdS)  
black holes (BHs) in particular exhibit rich thermal properties that differ qualitatively from those of their classical flat counterparts, allowing us to explore the stability and the phase structure of these configurations.

Within the framework of the gauge/gravity correspondence, the studies of PTs in strongly coupled quantum field theories have emerged as a subject of considerable importance \cite{Maldacena:1997re,Gubser:1998bc,Witten:1998qj}. A landmark result in this direction was established by Hawking and Page \cite{Hawking-Page}, who identified a critical temperature $T_c$ at which a PT occurs between thermal AdS spacetime and spherically symmetric AdS BHs. Above this critical temperature, BHs become thermodynamically favored, while below it, thermal AdS dominates. In the dual gauge theory description, this gravitational PT admits a natural interpretation as the confinement/deconfinement transition \cite{Witten:1998qj,Witten:1998zw}.

The inclusion of gauge fields naturally motivates the study of charged BHs. In this context, Chamblin et al. \cite{Chamblin:1999tk} carried out a detailed analysis of the phase structure of spherically symmetric charged BHs in both the canonical and grand-canonical ensembles, showing that the electric potential plays a central role in shaping the thermodynamic landscape. At sufficiently large potential, a single BH branch emerges as the thermodynamically preferred configuration in the canonical ensemble. Moreover, within the framework of extended BH thermodynamics, where the cosmological constant 
$\Lambda$ is interpreted as the dynamical pressure of the system \cite{Kubiznak:2016qmn,Mann:2025xrb}, BHs of fixed charge exhibit phase behavior analogous to that of Van der Waals fluids, including first-order phase transitions and critical phenomena \cite{Kubiznak:2012wp}.  The Hawking-Page PT likewise
can be interpreted as a liquid-solid transition 
\cite{Kubiznak:2014zwa}. 
The striking parallels between these gravitational phenomena and classical processes in chemical thermodynamics have motivated a molecular interpretation of the microscopic degrees of freedom responsible for BH entropy \cite{Wei:2019uqg}, a perspective that has evolved into the field now commonly known as BH chemistry \cite{Kubiznak:2016qmn,Mann:2025xrb}.

{An interesting situation arises for  BHs with planar horizons \cite{Lemos:1994xp,Vanzo:1997gw,Aminneborg:1997pz,Banados:1998dc,Mann:1997iz,Birmingham:1998nr} in that they have a counterpart soliton solution whose energy is lower than pure AdS spacetime 
  \cite{Horowitz:1998ha}.} To ensure that the thermodynamic quantities are well defined, at least one of the spatial directions must be compactified. This compactification introduces a negative Casimir energy and leads to the emergence of the AdS soliton spacetime, obtained via a double analytic continuation. Within this framework, the PT between a planar AdS BH and its AdS soliton counterpart can be studied by treating the soliton as the relevant thermal background \cite{Surya:2001vj}. Following the interpretation put forward in \cite{Witten:1998qj,Witten:1998zw}, this transition admits a natural dual description: the deconfinement phase corresponds to large, cold planar BHs, while the confinement phase is realized by small, hot AdS solitons \cite{Surya:2001vj}. 

More recently, increasing attention has been devoted to the study of PTs involving competing gravitational configurations, such as BHs and AdS solitons, particularly in the presence of nontrivial matter sources. In these scenarios, the interplay between electric and magnetic sectors can give rise to a rich structure of coexistence phases. For instance, in five dimensions, a phase transition between electrically charged Ricci-flat BHs and the corresponding AdS soliton was analyzed in \cite{Banerjee:2007by}. Extending this line of inquiry, \cite{Anabalon:2019tcy,Mohanty:2026hfg} studied the transition between hairy AdS BHs and solitons, where matter sources are explicitly incorporated into the field equations . More recently, PTs have been identified in the standard four-dimensional Einstein–Maxwell model with a negative cosmological constant $\Lambda$, involving magnetically charged planar solitons and electrically charged BHs \cite{Anabalon:2022ksf}.  In these scenarios, the occurrence of PTs depends sensitively on the electric potential, magnetic flux, and temperature within the grand-canonical ensemble, thereby enriching the phase structure of the theory.

{Nonlinear electrodynamics (NLE) can significantly enhance the thermodynamic structure of BHs 
\cite{Gunasekaran:2012dq,Gao:2021kvr,Gao:2025plm}.  The general form of the action for NLE coupled to Einstein gravity is 
\begin{equation}\label{eq:action-gao}
    S = \frac{1}{2\kappa} \int_{\mathcal{M}} d^4 x \ \sqrt{-g} \left( R - 2 \Lambda\right)-S_M=S_g-S_M,
\end{equation}
with the matter action expressed as
\begin{eqnarray}\label{eq:NLE-Gao}
     S_M&=&\frac{1}{2\kappa} \int_{\mathcal{M}} d^4x \sqrt{-g} L_M=\frac{1}{2\kappa} \int_{\mathcal{M}} d^4x \sqrt{-g}\,L (F)\nonumber\\
     &=&\frac{1}{2\kappa} \int_{\mathcal{M}} d^4x \sqrt{-g} \left(\sum_{k=1}^{+\infty} \alpha_k {F}^{k}\right),
\end{eqnarray}
leading to multicritical phenomena in which multiple phases can coexist at a particular temperature and pressure 
\cite{Tavakoli:2022kmo}.  Recently, triple points for BHs and AdS solitons in NLE were found
\cite{Quijada:2023fkc}.} However, a systematic framework for generating higher-order multicritical points in four-dimensional gravity has remained unexplored. In \eqref{eq:NLE-Gao}, ${F}=F_{\mu \nu} F^{\mu \nu}$ with $F_{\mu \nu}=\partial_{\mu} A_{\nu}-\partial_{\nu} A_{\mu}$, the $\alpha_i$'s denote dimensionful coupling constants, while $A_{\mu}$ represents the $U(1)$ Maxwell field. It is important to emphasize that the number of event horizons depends on the number of free coupling constants $\alpha_i$'s. For instance, BH solutions with three horizons arise when $\alpha_2$ is chosen freely (with $\alpha_1$ fixed to unity), while configurations with four horizons require two independent parameters, $\alpha_2$ and $\alpha_3$ (where as before $\alpha_1=1$); the remaining coupling constants are then determined in terms of these \cite{Gao:2021kvr}.

To address  generating higher-order multicritical points, we show in this paper that for a class of NLE models, originally formulated in \cite{Plebanski:1968,Salazar:1987}, the dynamics can be described by the action:
\begin{eqnarray}\label{eq:NLE}
S_{P} &=&\frac{1}{2\kappa} \int_{\mathcal{M}} d^4x \sqrt{-g} L_P\nonumber\\
&=&\frac{1}{2\kappa} \int_{\mathcal{M}} d^4x \sqrt{-g} \big(2 F_{\mu \nu} P^{\mu \nu} - \mathcal{H}(P)\big) \ ,
\end{eqnarray}
together with a polynomial expression of the structural function of the form:
\begin{eqnarray}\label{structural-function}
\mathcal{H}({P})=\sum_{i=1}^{n} \beta_i {P}^{i},
\end{eqnarray}
where the invariant $P$ is given by $P=P_{\mu \nu} P^{\mu \nu}$, with $P_{\mu \nu}$ the Pleba\'nski tensor. It is important to note that the matter source (\ref{eq:NLE}) has proven to be a useful tool in finding regular \cite{Ayon-Beato:1998hmi,Ayon-Beato:2000mjt,Ayon-Beato:1999kuh}, rotating \cite{DiazGarcia:2022jpc,Garcia-Diaz:2021bao,Ayon-Beato:2022dwg,Bravo-Gaete:2025xya}, AdS \cite{Alvarez:2022upr,Lin:2024ubg,Bravo-Gaete:2025fwx}, charged and hairy \cite{Bravo-Gaete:2022mnr,Hyun:2019gfz,Churilova:2019sah}  BHs.

As a first goal, we establish a mapping between the standard NLE Lagrangian (\ref{eq:NLE-Gao}) and the Pleba\'nski formulation (\ref{eq:NLE})-(\ref{structural-function}). This construction reproduces the complete set of charged BHs reported in \cite{Gao:2021kvr}, independently of the choice of base manifold, where the number of free parameters $\alpha_i$ (and, in consequence, the number of horizons) is directly tied to the polynomial degree $n \in \mathbb{N}$.  Secondly, through this mapping, we construct new families of charged AdS BH and soliton solutions, thereby extending the existence of the triple points previously identified in \cite{Quijada:2023fkc} to multi-point phases. These correspond to regions of the thermodynamic parameter space
where several locally stable gravitational configurations coexist with identical Gibbs free energy. 

We demonstrate that this construction provides a controlled mechanism for generating multicritical behavior, where the degree of the polynomial (\ref{structural-function}) directly determines the number of coexisting phases, where the coexistence of multiple magnetically charged solitons, with numerous swallowtail transitions, and an electrically charged BH is allowed. In contrast, the BH branch does not display swallowtail behavior, and it does not allow multiple electrically charged BHs to coexist with
a magnetically charged soliton. \\

Our paper is organized as follows: In Section \ref{relation}, we will show the explicit relation between the matter sources (\ref{eq:NLE-Gao}) and (\ref{eq:NLE}) with $\mathcal{H}(P)$ given by (\ref{structural-function}), recovering the cases reported in Ref. \cite{Gao:2021kvr}. In Section \ref{sol-termo}, we construct electrically charged AdS BH and magnetically charged soliton solutions, obtaining their thermodynamic quantities. In Section \ref{PT}, we analyze the PTs between these configurations and demonstrate that multiple magnetically charged solitons can coexist with a single electrically charged AdS BH, whereas the converse situation (multiple electrically charged BHs coexisting with a magnetically charged soliton) does not occur. Finally, in Section \ref{conclusions} we present our conclusions and perspectives.

\section{Relation between $L_M$ and $L_P$}\label{relation}

As discussed in the introduction, a central goal of this investigation is to establish a systematic relation between NLE theories formulated in terms of the Maxwell invariant $F$ (\ref{eq:NLE-Gao}), and those expressed in the Pleba\'nski framework (\ref{eq:NLE})-(\ref{structural-function}),  where the polynomial degree $n$ dictates the number of free parameters $\alpha_i$ together with the resulting number of horizons. This mapping provides a constructive bridge between the two formulations and plays a key role in generating multicritical behavior. To this end, the equations of motion for the matter action (\ref{eq:NLE}) read:  
\begin{eqnarray}
     T^{P}_{\mu \nu}&=& 2 \mathcal{H}_P P_{\mu \alpha} P_{\nu}^{\ \alpha} - \frac{1}{2} g_{\mu \nu} \left( 2 P \mathcal{H}_P - \mathcal H \right), \label{eq:Tmunu}\\
    \nabla_\mu P^{\mu \nu} &=& 0, \label{eq:Pmunu} \\
    F_{\mu \nu} &=& \mathcal{H}_P P_{\mu \nu}, \label{eq:constitutive}
\end{eqnarray}
where $\mathcal{H}_P$ denotes $\partial \mathcal{H}/\partial P$.  This last relation encodes the nonlinear response of the electromagnetic field and is responsible for the nontrivial mapping between the invariants $F$ and $P$. \\

The  asymptotically AdS metric ansatz is taken to be:
\begin{eqnarray}\label{eq:metricbh}
    ds_b^2 &=&-U_b(r) dt_b^2 + \frac{dr^2}{U_b(r)} + \frac{r^2}{\ell^2} \left(  d\varphi^2_b + dz^2 \right),
\end{eqnarray}
together with a potential:
\begin{eqnarray}\label{eq:potential}
    A_b &=& A_t (r) dt_b + A_\varphi d\varphi_b,
\end{eqnarray}
where $0 \leq \varphi_b \leq \eta_b$, $0 < z < L$ and $A_\varphi$ are arbitrary constants. \\

From eq. \eqref{eq:constitutive}, the invariants ${P}$ and ${F}$ can be expressed as:
$$F={F_{\mu \nu}}{F^{\mu \nu}}=2F_{tr} F^{tr}=\big(\mathcal{H}_{P}\big)^{2}\,P.$$
Substituting the explicit form of the structural function (\ref{structural-function}), we obtain:
$$\mathcal{H}_{P}=\sum_{i=1}^{n} i \beta_{i}{P}^{i-1},$$
which leads to
\begin{eqnarray}\label{eq:EOM}
{F}   
= \left(\sum_{i=1}^{n} i \beta_{i}{P}^{i-1}\right)^{2} {P}.
\end{eqnarray}

This expression provides a nonlinear redefinition of the invariant $F$ in terms of $P$, whose structure depends explicitly on the polynomial degree $n$. On the other hand, from the definition of the Pleba\'nski Lagrangian, one finds
\begin{eqnarray}
\sum_{i=1}^{n} (2i-1)\beta_{i} {P}^{i} 
&=& 2 \mathcal{H}_{P} {P} - \mathcal{H}({P}) \label{mastereq1}\\
&=& 2 F_{\mu\nu}P^{\mu\nu} - \mathcal{H}({P})= \sum_{k=1}^{+\infty} \alpha_{k}\,{F}^{k} \nonumber\\ 
&=& \sum_{k=1}^{+\infty} \alpha_{k}\left(\sum_{i=1}^{n} i\beta_{i} {P}^{i-1}\right)^{2k}\! {P}^{k}. \nonumber
\end{eqnarray}  
This identity establishes a nontrivial algebraic relation between the coefficients $\alpha_k$ and $\beta_i$, which can be made explicit using combinatorial methods. Expanding the power using the multinomial theorem \cite{Russel:2003}, for a non-negative integer $n$:  
\begin{eqnarray}
\left(\sum_{i=1}^{n} i\beta_{i} {P}^{i-1}\right)^{2k}  
&=& \sum_{\{k_1,\dots,k_n\}}
\frac{(2k)!}{k_1! \cdots k_n!}
\prod_{j=1}^{n} \big(j\beta_{j} {P}^{\,j-1}\big)^{k_j}, \nonumber
\end{eqnarray}  
where the sum is taken over all non-negative integer solutions of  \begin{eqnarray}\label{eq:sum}
\sum_{j=1}^{n} k_{j} = 2k.
\end{eqnarray} 
Substituting back, one obtains a polynomial expansion in powers of 
$P$, which can be matched term by term with $\sum_{i=1}^{n} (2i-1)\beta_{i} {P}^{i}$ as follows:
\begin{eqnarray}\label{mastereq2}
\sum_{k=1}^{+\infty}  \alpha_{k}  
\Bigg[ \sum_{\{k_1,\dots,k_n\}}
\frac{(2k)!}{k_1! \cdots k_n!}  
\prod_{j=1}^{n}\big(j\beta_{j}\big)^{k_j} \Bigg]  
\nonumber\\
= \begin{cases}  
(2i-1)\beta_{i}, & \mbox{ for } \quad 1 \leq i \leq n, \\  
0, & \mbox{ for } \quad i > n,  
\end{cases}
\end{eqnarray}  
subject to the additional constraint  
\begin{eqnarray}\label{eq:sum2}
\sum_{j=1}^{n-1} j\,k_{j+1}= i-k, \qquad \mbox{ for } \quad 1 \leq i \leq n, \quad i \geq k.
\end{eqnarray} 

This mapping shows that a polynomial structural function of degree $n$ (\ref{structural-function}) in the Pleba\'nski formulation corresponds to an infinite series in the Maxwell invariant (\ref{eq:NLE-Gao}), whose coefficients are completely determined by the finite set of couplings $\beta_i$. This establishes that the polynomial degree controls the number of independent parameters and, consequently, the number of horizons and thermodynamic branches that may arise in the corresponding solutions.\\

As a pedagogical illustration, consider the quadratic case $n=2$, where we obtain
$$\beta_1 = \frac{1}{\alpha_1}, \quad  
\beta_2 = -\frac{\alpha_2}{\alpha_1^{4}},$$
while the higher-order coefficients are related as: 
\begin{eqnarray}\label{eq:case1}
 \alpha_3 = \frac{4\alpha_2^{2}}{\alpha_1},\quad
\alpha_4 = \frac{24\alpha_2^{3}}{\alpha_1^{2}},\quad 
\alpha_5 = \frac{176\alpha_2^{4}}{\alpha_1^{3}}, \quad \cdots
\end{eqnarray}  
while for the cubic situation $n=3$, we obtain  
\begin{eqnarray*}
\beta_1 &=& \frac{1}{\alpha_1}, \quad  
\beta_2 = -\frac{\alpha_2}{\alpha_1^{4}}, \quad  
\beta_3 = \frac{4\alpha_2^{2}-\alpha_1\alpha_3}{\alpha_1^{7}}, \label{eq:case2}
\end{eqnarray*}
with higher-order coefficients
\begin{eqnarray}
\alpha_4 &=& \frac{12\alpha_2(\alpha_1\alpha_3-2\alpha_2^{2})}{\alpha_1^{2}}, \nonumber\\
\alpha_5 &=& \frac{60\alpha_1\alpha_2^{2}\alpha_3 - 208\alpha_2^{4} + 9\alpha_1^{2}\alpha_3^{2}}{\alpha_1^{3}}, \nonumber\\
\alpha_6 &=& \frac{26\alpha_2(9\alpha_1^{2}\alpha_3^{2}-16\alpha_1\alpha_2^{2}\alpha_3 - 24\alpha_2^{4} )}{\alpha_1^{4}}, \quad \cdots \nonumber
\end{eqnarray}  
recovering the relations reported in \cite{Gao:2021kvr} for the specific cases of BH solutions for 3-horizon and 4-horizon (with the choice $\alpha_1 = 1$), and explicitly demonstrate how increasing the polynomial degree introduces additional nonlinear interactions. \\

The above construction provides a systematic and controllable framework for generating increasingly rich gravitational solutions. In particular, the polynomial degree $n$ acts as an organizing parameter that determines both the complexity of the horizon structure and the richness of the thermodynamic phase space. As we will show in the following sections, this structure is directly responsible for the emergence of multicritical behavior, where multiple gravitational phases can coexist. \\

\textcolor{black}{We pause to comment on the choice of the Pleba\'nski
frame. The formulations (\ref{eq:NLE-Gao})-(\ref{structural-function}) are related by a Legendre-type
transformation, $L(F)=2F_{\mu\nu}P^{\mu\nu}-\mathcal{H}(P)$, together with
the constitutive relation (\ref{eq:constitutive}); on any domain where the
latter is invertible ($\mathcal{H}_P\neq 0$), and the two descriptions are
physically equivalent on-shell. The Pleba\'nski frame is nevertheless
singled out for practical and structural reasons. First, for the static
electric (and, after the double Wick rotation, magnetic) configurations that will be considered here, eq. (\ref{eq:Pmunu}) integrates immediately, so that the
invariant $P$ becomes a universal function of the radial coordinate and the
field equations reduce to quadratures for any polynomial degree $n$. On the other hand,  in the
$F$-frame, the same problem admits an implicit or order-by-order analysis, as was performed in \cite{Gao:2025plm,Gao:2021kvr}, 
obscuring the global structure of the Gibbs free energy on which the
multicriticality analysis relies. Second, as the mapping
(\ref{mastereq1})-(\ref{eq:sum2}) makes explicit, a structural function
truncated at a finite polynomial order $n$ corresponds to a
{non-polynomial} Lagrangian $L(F)$, defined as an infinite power series
in $F$ whose coefficients $\alpha_k$ are completely fixed by the finite set of $\beta_i$'s.  The Pleba\'nski frame thus provides a
resummed, closed-form description of the theory with a finite number of couplings $\beta_i$'s. Finally, the polynomial form of $\mathcal{H}(P)$ is not
essential for the existence of swallowtail structures, as will be discussed in Section \ref{PT}, where it persists under small deformations of the couplings and
would also arise for suitable non-polynomial structural functions. Instead, as a novelty, polynomiality provides the organizing principle of our construction, where the degree $n$ directly controls the number of admissible extrema of the thermodynamic functions and, in consequence, the maximal number of
coexisting phases. }

Building on the above results \textcolor{black}{and comments}, we now focus on the toy model introduced in (\ref{eq:action-gao}), where matter contribution $S_M$ is substituted by the Pleba\'nski-type action $S_P$ (see eq.~(\ref{eq:NLE})), constructed with the polynomial structural function (\ref{structural-function}).

\section{Black hole and soliton thermodynamics }\label{sol-termo}
We consider an action given by the general form of the Pleba\'nski NLE (\ref{eq:NLE}) minimally coupled to the four-dimensional Einstein-Hilbert action together with a cosmological constant; namely, $S=S_g-S_P$, where the Einstein equations with respect to the metric $g_{\mu \nu}$ read:
\begin{eqnarray}\label{eq:einstein}
    R_{\mu \nu} - \frac{1}{2} g_{\mu \nu} R + \Lambda g_{\mu \nu} &=& T^{P}_{\mu \nu},
\end{eqnarray}
with $T^{P}_{\mu \nu}$ given previously in eq. (\ref{eq:Tmunu}). \\

The equations of motion for the gauge field and the Pleba\'nski tensor are given in eqs. (\ref{eq:Pmunu})-(\ref{eq:constitutive}), while the metric ansatz and the potential are specified in  eqs. (\ref{eq:metricbh})-(\ref{eq:potential}). As before, we restrict our attention to purely electric configurations. From eq.~(\ref{eq:Pmunu}), the antisymmetric Pleba\'nski tensor $P_{\mu \nu}$ takes the form
\begin{eqnarray}
    P_{\mu \nu} &=& 2 \delta^t_{[\mu} \delta^r_{\nu]}  \frac{Q}{r^2},
\end{eqnarray}
where $Q$ denotes an integration constant related to the electric charge, and the invariant $P$ is given by $$P=-\frac{2 Q^2}{r^{4}}.$$
Using this ansatz, from the field equations (\ref{eq:Tmunu})-(\ref{eq:constitutive}) together with (\ref{eq:einstein}), we arrive at the following set of equations:
\begin{eqnarray}
    A_t' -\frac{r^2}{2Q} \sum_{i=1}^{n} \beta_i \ i  \left(\frac{-2Q^2}{r^4} \right)^i &=& 0, \\
    rU_b' + U_b - \frac{3 r^2}{\ell^2} - \frac{r^2}{2} \sum_{i=1}^{n} \beta_i \left( \frac{-2 Q^2}{r^4} \right)^i &=& 0,
\end{eqnarray}
where the prime denotes the derivative with respect to the radial coordinate $r$. 

Integrating these equations leads to
\begin{eqnarray}
    A_t(r) &=& \sum_{i=1}^{n} \frac{i (-2)^{i-1} \beta_i Q^{2i-1}}{(4i-3) r^{4i-3}} \nonumber\\
    &-& \sum_{i=1}^{n} \frac{i (-2)^{i-1} \beta_i Q^{2i-1}}{(4i-3) r_{+}^{4i-3}}, \\
    U_b(r) &=& {\frac{r^2}{\ell^2} - \frac{2 M}{r} - \frac{1}{2} \sum_{i=1}^{n} \frac{\beta_i (-2Q^2)^{i}}{(4i-3) r^{4i-2}}},
\end{eqnarray}
with $M$ denoting an integration constant associated with the BH mass and $r=r_+$ represents the location of the (outer) event horizon. \\

These solutions generalize charged planar AdS BHs by incorporating NLE corrections controlled by the coefficients $\beta_i$'s, where the polynomial degree $n$ determines the number of independent contributions to the metric function, which in turn influences the horizon structure. \\

On the other hand,  a double Wick rotation  given by $$t_b \rightarrow i \varphi_b \qquad \varphi_b \rightarrow i t_b,$$  and applied to the metric Ansatz (\ref{eq:metricbh})  yields  the following magnetically charged soliton metric, gauge potential, and Pleba\'nski tensor as follows:
\begin{eqnarray}
    ds_s^2 &=&- \frac{r^2}{\ell^2} dt_s^2 + \frac{dr^2}{U_s(r)} +  U_s(r) d\varphi^2_s + \frac{r^2}{\ell^2} dz^2, \label{eq:metricsol} \\
    A_s &=& A_t dt_s + A_\varphi (r) d\varphi_s  \ , \\
    P_{\mu \nu} &=& 2 \delta^\varphi_{[\mu} \delta^r_{\nu]}  \frac{q}{r^2} \ .
\end{eqnarray}
where $0 \leq \varphi_s \leq \eta_s$, $0 < z < L$, $A_t$ is an arbitrary constant, and now the invariant $P$ reads 
$$P=P_{\mu \nu}P^{\mu \nu}=\frac{2 q^2}{r^{4}},$$
where $q$ is another integration constant related now to the magnetic charge of the soliton. As before, through the equations of motion (\ref{eq:Tmunu})-(\ref{eq:constitutive}) and (\ref{eq:einstein}), we obtain:
\begin{eqnarray}
    A_\varphi' + \frac{r^2}{2q} \sum_{i=1}^{n} i \beta_i   \left(\frac{2 q^2}{r^4} \right)^i &=& 0, \\
    rU_s' + U_s - \frac{3 r^2}{\ell^2} - \frac{r^2}{2} \sum_{i=1}^{n}\beta_i \left( \frac{2 q^2}{r^4} \right)^i &=& 0,
\end{eqnarray}
and the solitonic configuration takes the form
\begin{eqnarray}
    A_\varphi(r) &=& \sum_{i=1}^{n} \frac{2^{i-1}  i \beta_i q^{2i-1}}{(4i-3) r^{4i-3}} - \sum_{i=1}^{n} \frac{2^{i-1} i \beta_i q^{2i-1}}{(4i-3) r_{0}^{4i-3}}, \\
    U_s(r) &=& \frac{r^2}{\ell^2} - \frac{2 \mu}{r} - \frac{1}{2} \sum_{i=1}^{n} \frac{2^i \beta_i q^{2i}}{(4i-3) r^{4i-2}}.
\end{eqnarray}
The soliton geometry is regular provided that $U_s(r_0)=0$ for some $r_0>0$, where $r_0$ defines the minimum value of the radial coordinate. The geometry then closes off smoothly at $r=r_0$, while $\mu$ is an integration constant related to the mass. The absence of conical singularities imposes the constraint
\begin{equation}
 g_s(r_0,q)=\left( \frac{3 r_0}{4 \pi \ell^2} + \frac{1}{8 \pi} \sum_{i=1}^{n} \frac{\beta_i \left(2q^{2}\right)^i}{r_0^{4i-1}} \right)-\frac{1}{\eta_s}=0,\label{eq:g-condition}
\end{equation}
which fixes the period $\eta_s$ of the soliton coordinate $\varphi_s$, and introduces a nontrivial relation between the parameters $r_0$ and $q$. \\
Concerning the thermodynamics of both configurations, the on-shell Euclidean action is given by
\begin{equation}\label{eq:euclidean-action}
    I_\mathrm{Euc}=-I_\mathrm{bulk} - I_\mathrm{GH} + I_\mathrm{ct}\ ,
\end{equation}
where 
$$I_\mathrm{bulk}=\frac{1}{2\kappa} \int_{\mathcal{M}} d^4 x \ \sqrt{g} \left( R - 2 \Lambda-L_P\right),$$ is the Euclidean bulk action and  
$$I_\mathrm{GH}=\frac{1}{\kappa} \int_{\partial \mathcal{M}} d^3 x \ \sqrt{h}\,K,$$ denotes the Gibbons-Hawking boundary term, where $K$ is the trace of the extrinsic curvature of the boundary and $h_{ij}$ is its induced metric, while that 
$$I_\mathrm{ct}=\frac{1}{\kappa} \int_{\partial \mathcal{M}} d^3 x \ \sqrt{h} \left(\frac{2}{\ell}-\frac{\ell}{2} \mathcal{R}(h)\right),$$
represents an appropriate local counterterm, with $\mathcal{R}$ denoting the Ricci scalar of the boundary metric $h_{ij}$. From the Euclidean action (\ref{eq:euclidean-action}), we obtain that the free energies of the BH as well as the soliton, denoted by $G_b$ and $G_s$, are given by
\begin{eqnarray}
    G_b &=& - \frac{L \eta_b}{\kappa \ell^2} M\nonumber\\
    &=&{- \frac{L \eta_b}{\kappa \ell^2} \left( \frac{r_+^3}{2 \ell^2} - \frac{1}{4} \sum_{i=1}^{n} \frac{ \beta_i(-2 Q^2)^i}{(4i-3) r_+^{4i-3}}\right)},\label{eq:Gb}
\\
    G_s &=& - \frac{L \eta_s}{\kappa \ell^2} \mu \nonumber\\
    &=&{- \frac{L \eta_s}{\kappa \ell^2} \left(\frac{r_0^3}{2 \ell^2} - \frac{1}{4} \sum_{i=1}^{n} \frac{\beta_i {(2q^2)^{i}}}{(4i-3) r_0^{4i-3}}\right). \label{eq:Gs}}
\end{eqnarray}
Finally, the Hawking temperature $T_b$, the electric potential $\phi_b$ for the BH, and the magnetic flux $\Phi_s$ are expressed as:
\begin{eqnarray}
    T_b &=& \frac{3 r_+}{4 \pi \ell^2} + \frac{1}{8 \pi} \sum_{i=1}^{n} \frac{ \beta_i (-2Q^2)^i}{r_+^{4i-1}}, \label{eq:Tb}\\
    \phi_b &=&  {\sum_{i=1}^{n} \frac{(-2)^{i-1} i \beta_i  Q^{2i-1}}{(4i-3) r_{+}^{4i-3}}}, \label{eq:phib} \\
    \Phi_s &=&  {\left( \sum_{i=1}^{n} \frac{2^{i-1} i \beta_i  q^{2i-1}}{(4i-3) r_{0}^{4i-3}} \right) \eta_s \ .}\label{eq:Phis} 
\end{eqnarray}
An important feature of this construction is that the NLE parameters $\beta_i$  appear in all thermodynamic quantities, thereby directly influencing the phase structure. In particular, the dependence on higher powers of the charges introduces additional extrema in the thermodynamic potentials, which will be shown to generate multicritical behavior in the soliton sector. This establishes a direct link between the microscopic structure of the matter sector and the system's macroscopic thermodynamic properties. \\

With these ingredients at hand, we are now in a position to analyze the PT and the emergence of multicritical points.

\section{Phase Transitions }\label{PT}

We now turn to the analysis of PTs between the electrically charged BH and the magnetically charged soliton configurations constructed in the previous section. \\

Our goal is to determine the thermodynamically preferred phase and to identify the conditions under which multiple configurations can coexist. The analysis is performed in the grand-canonical ensemble, where the temperature and the electromagnetic variables 
$(T,\phi_b,\Phi_s)$ are held fixed at the boundary. \\

To study PTs between the solutions, we first consider the Euclidean planar BH and soliton, which are obtained from the line elements (\ref{eq:metricbh}) and (\ref{eq:metricsol}) via the coordinate transformations: 
$$t_b \to - i \tau_b , \qquad t_s \to - i \tau_s.$$
Here, both Euclidean coordinates are taken to be periodic ($0 \leq \tau_b \leq \beta_b$ and $0 \leq \tau_s \leq \beta_s$) with periods $\beta_b=1/T$, and $\beta_s$ respectively. Then, to match the asymptotic geometries at a finite radial cut-off surface $r=\rho$, we match the induced boundary metrics by equating the proper lengths of the Euclidean time circle and of the compact spatial circle. This ensures both geometries share the same boundary data, allowing a meaningful comparison of their Euclidean action, which is:
\begin{equation}
\sqrt{U_b(\rho)}\, \beta_b=\left(\frac{\rho}{\ell}\right) \beta_s,\qquad \left(\frac{\rho}{\ell}\right) \eta_b=\sqrt{U_s(\rho)}\, \eta_s,
\end{equation}
where at the limit  $\rho \rightarrow +\infty$, one finds that 
$$\beta_b = \beta_s = \beta, \qquad \eta_b = \eta_s = \eta.$$
This matching procedure ensures that both geometries are evaluated under identical thermal and geometric boundary conditions. In the grand-canonical ensemble, the boundary sources are held fixed. Consequently, a meaningful comparison between the competing phases requires the electric potential of the BH to coincide with that of the soliton, while the magnetic flux must be the same in both geometries. \\

We begin by analyzing the soliton sector, which will be shown to be responsible for the emergence of multicritical behavior. The key quantities are the magnetic flux 
$\Phi_s(r_0,q)$ (eq. (\ref{eq:Phis})) and the Gibbs free energy $G_s(\Phi_s)$ (eq. (\ref{eq:Gs})), both subject to the regularity constraint $g(r_0,q)=0$ (eq. \ref{eq:g-condition}).  From a mathematical perspective, the problem is to determine the extrema of $\Phi_s$ and $G_s$ subject to the constraint (\ref{eq:g-condition}). {This can be implemented via the obtention of $r_0^{\mbox{\tiny{ext}}}>0$ such that:
\begin{eqnarray*}
\frac{d \Phi_s}{ d r_0} \Bigg{|}_{r_0^{\mbox{\tiny{ext}}}}=0=\frac{d G_s}{d r_0} \Bigg{|}_{r_0^{\mbox{\tiny{ext}}}},\\
\left(\frac{d^2 \Phi_s}{d r_0^2} \Bigg{|}_{r_0^{\mbox{\tiny{ext}}}}\right)\left(\frac{d^2 G_s}{d r_0^2} \Bigg{|}_{r_0^{\mbox{\tiny{ext}}}}\right)>0,
\end{eqnarray*}
where all derivatives are total derivatives along the curve defined by (\ref{eq:g-condition}), with $q=q(r_0)$ determined implicitly.} 

It is important to distinguish between two levels of genericity in our construction. The appearance of swallowtail structures in $G_s(\Phi_s)$ is a generic phenomenon. Indeed, once the extrema of $\Phi_s$ and $G_s$ are present, their persistence under small deformations of the couplings $\beta_i$ follows from the implicit function theorem. Consequently, swallowtail behavior occurs throughout an open region of parameter space and does not require fine tuning. \\

\begin{figure*}[t]
\centering

\begin{subfigure}[c]{0.3\textwidth}
    \centering
    \includegraphics[width=\linewidth]{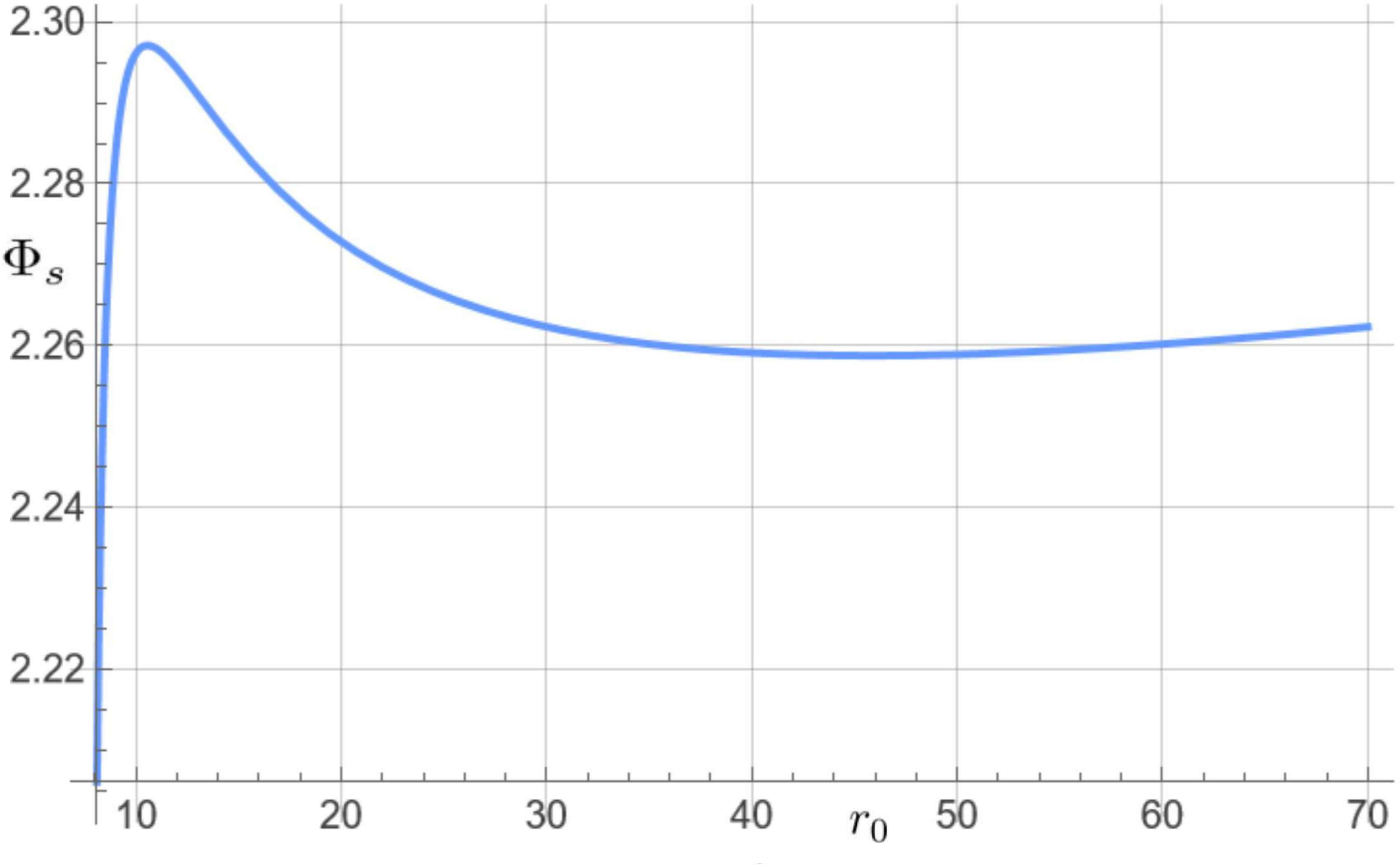}
    \caption{$\Phi_s$ as a function of $r_0$.}
\end{subfigure}
\hfill
\begin{subfigure}[c]{0.3\textwidth}
    \centering
    \includegraphics[width=\linewidth]{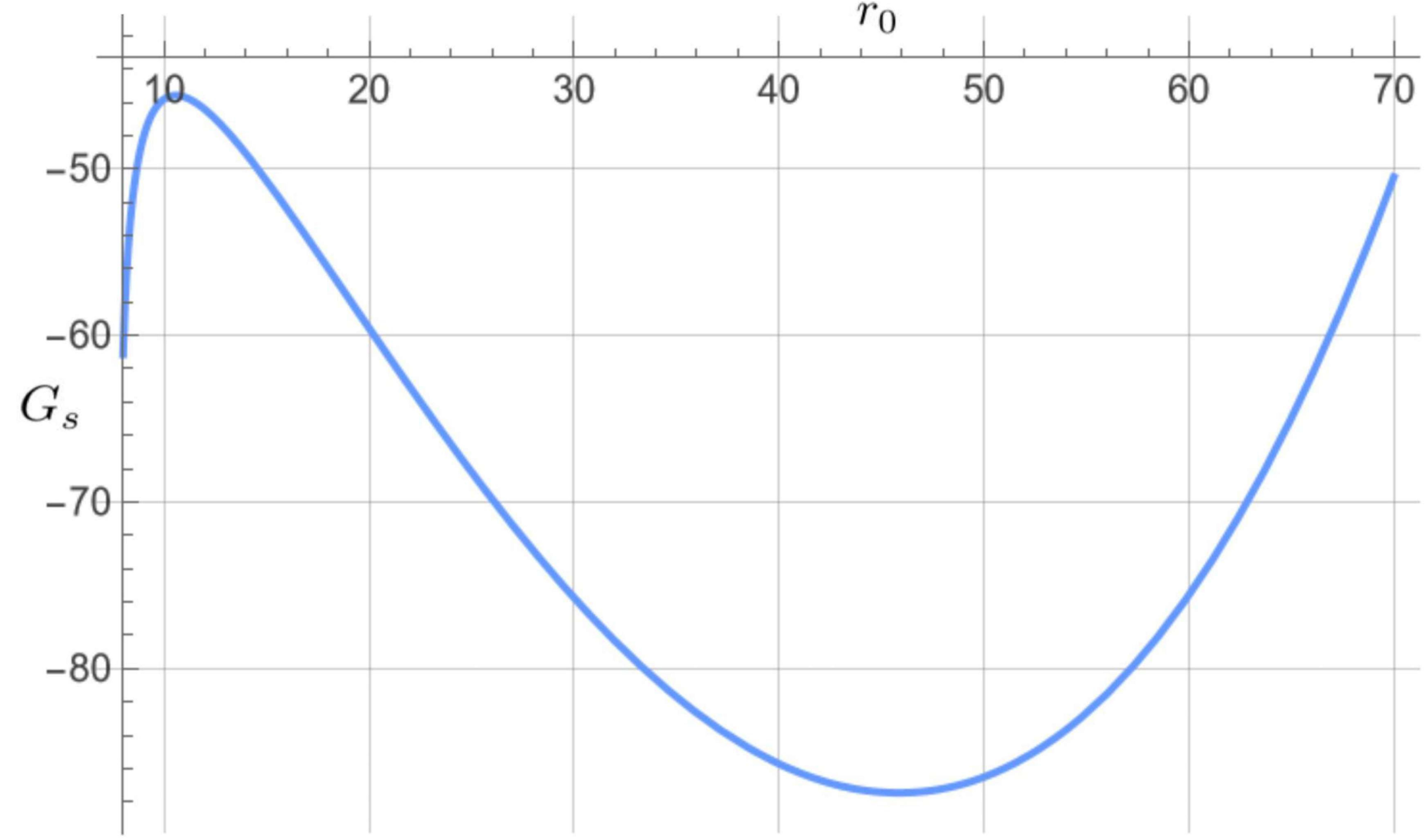}
    \caption{$G_s$ as a function of $r_0$.}
\end{subfigure}
\hfill
\begin{subfigure}[c]{0.3\textwidth}
    \centering
    \includegraphics[width=\linewidth]{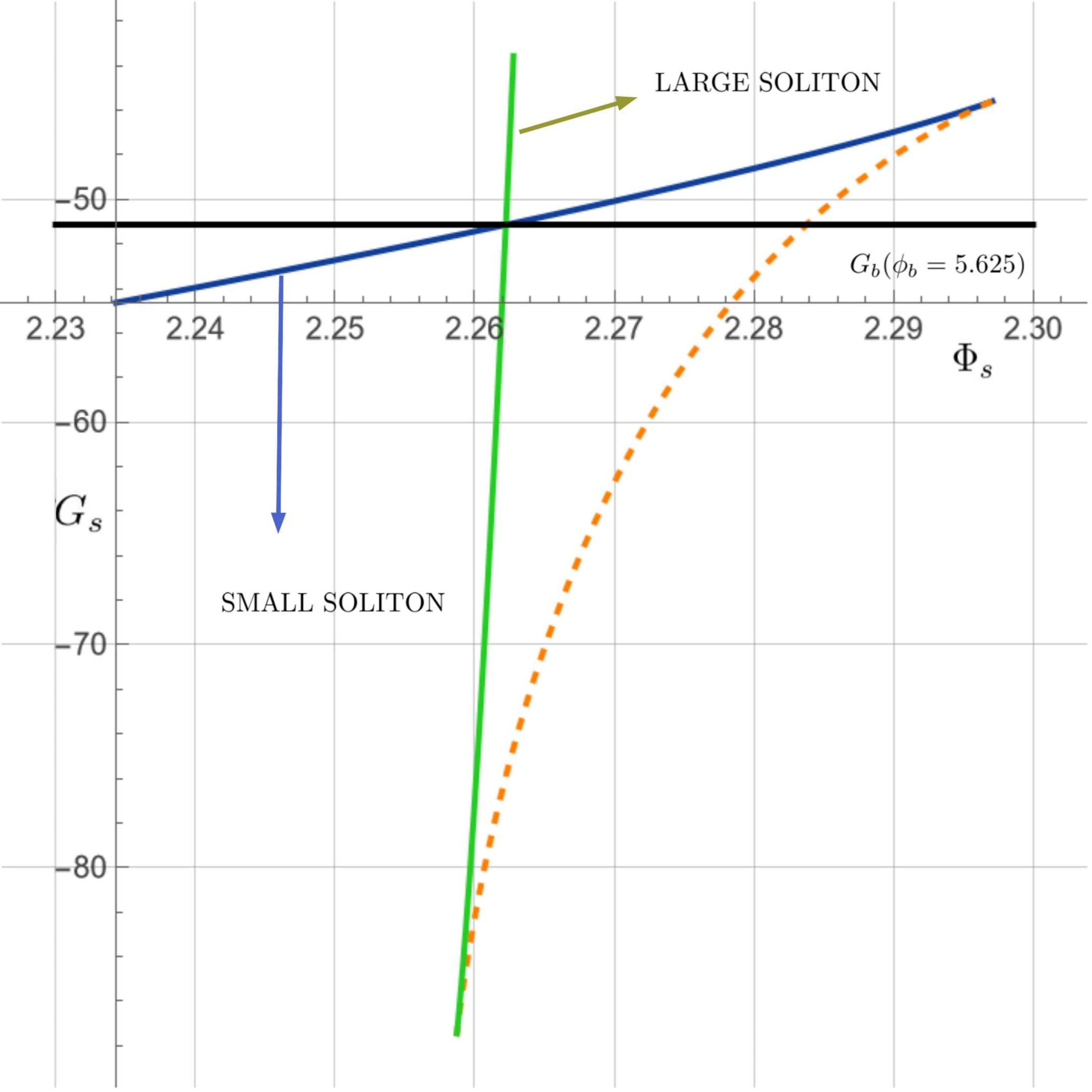}
    \caption{$G_s$ depending on $\Phi_s$.}
\end{subfigure}

\vspace{-0.2cm}

\caption{For $\beta_1 = 1$ and $ \beta_2 = -0.624$, we recover the previous result \cite{Quijada:2023fkc} of the coexistence of one BH and two locally stable solitons (the small soliton and large soliton represented through the blue and green curves, respectively). Here, $\eta=\kappa=L=0.5=\ell/2$, $T_b=0.6$, and the electric potential takes the value $\phi_b=5.625$.}
\label{fig1}
\end{figure*}

\begin{figure*}[t]
\centering

\begin{subfigure}[c]{0.30\textwidth}
    \centering
    \includegraphics[width=\linewidth]{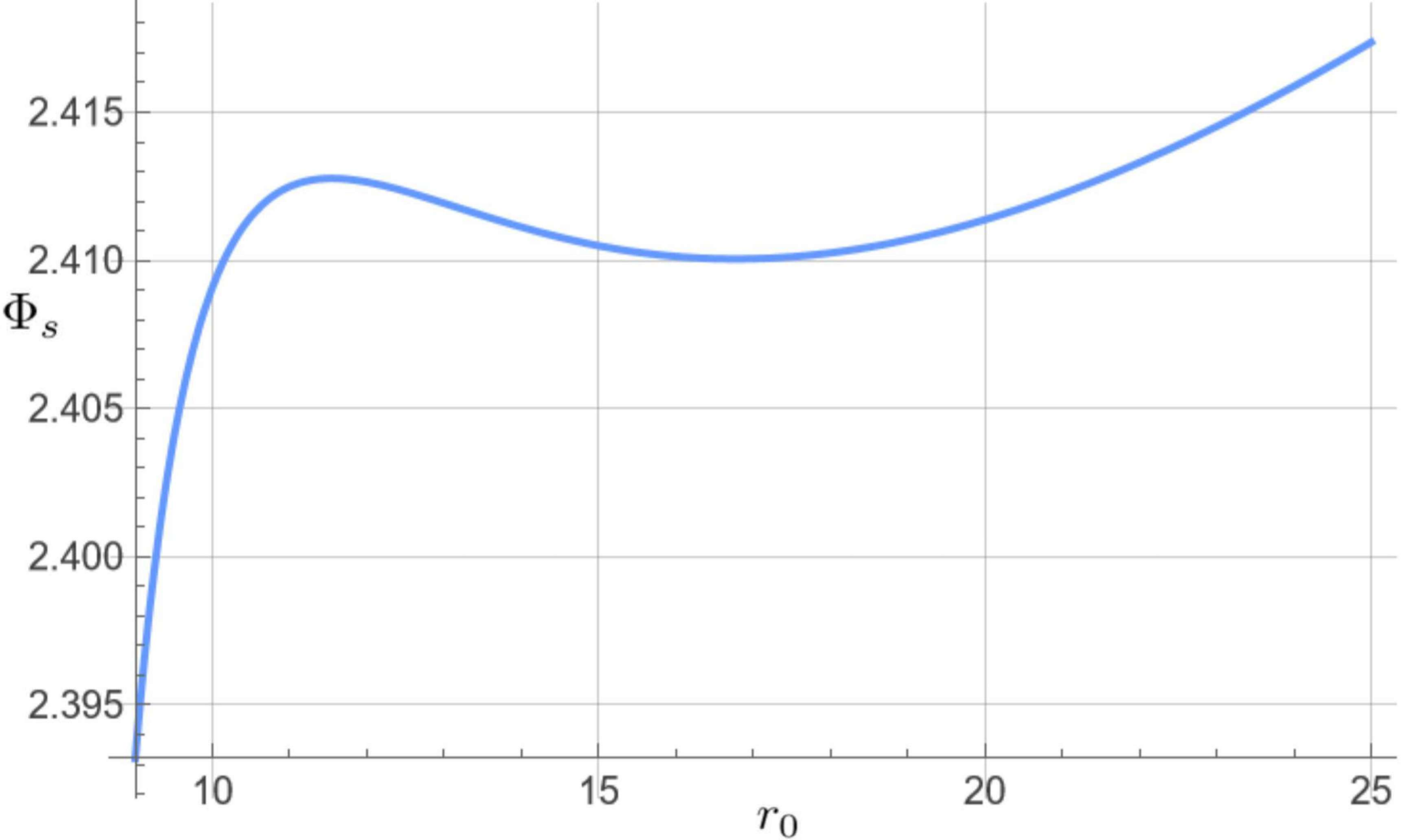}
    \caption{$\Phi_s$ as a function of $r_0$.}
\end{subfigure}
\hfill
\begin{subfigure}[c]{0.30\textwidth}
    \centering
    \includegraphics[width=\linewidth]{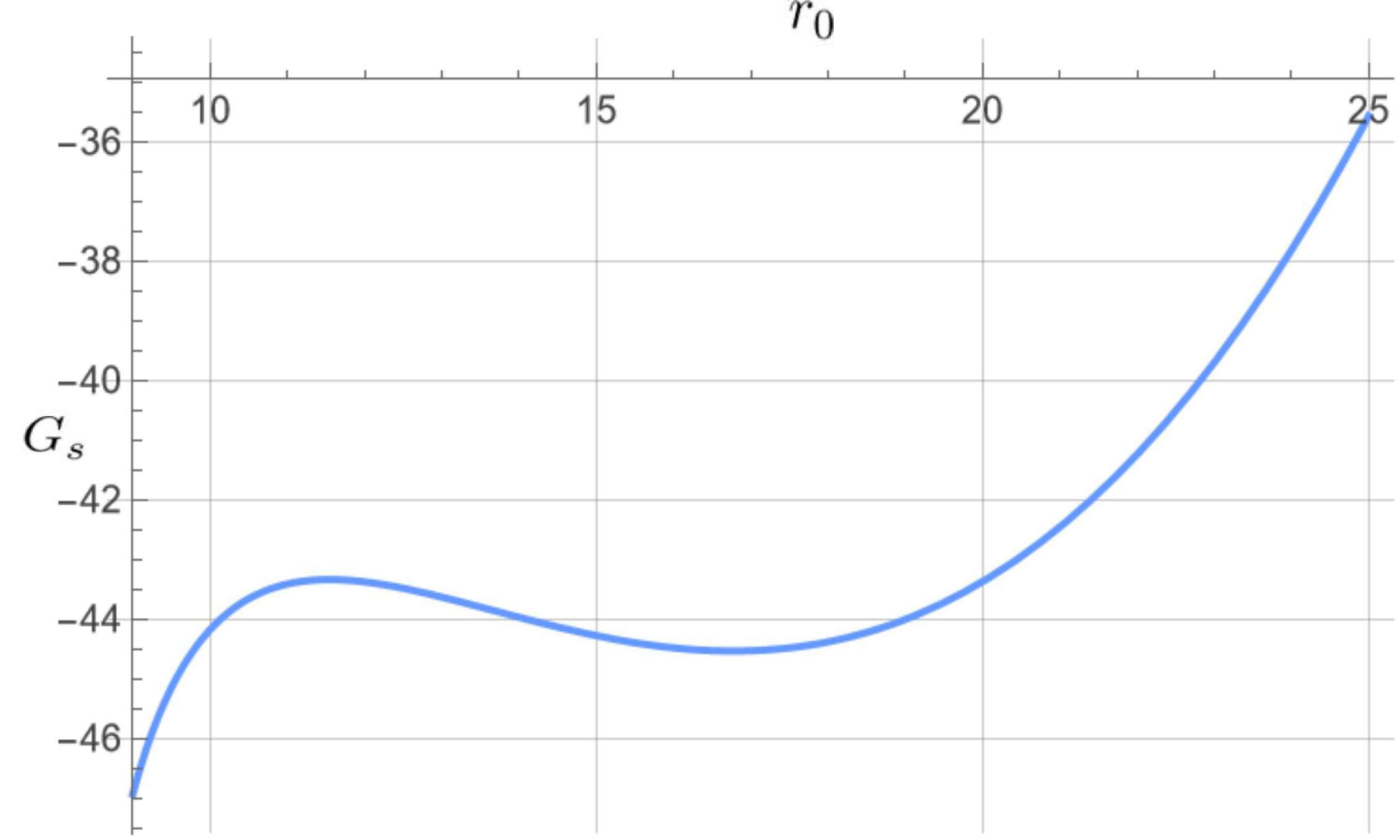}
    \caption{$G_s$ as a function of $r_0$.}
\end{subfigure}
\hfill
\begin{subfigure}[c]{0.30\textwidth}
    \centering
    \includegraphics[width=\linewidth]{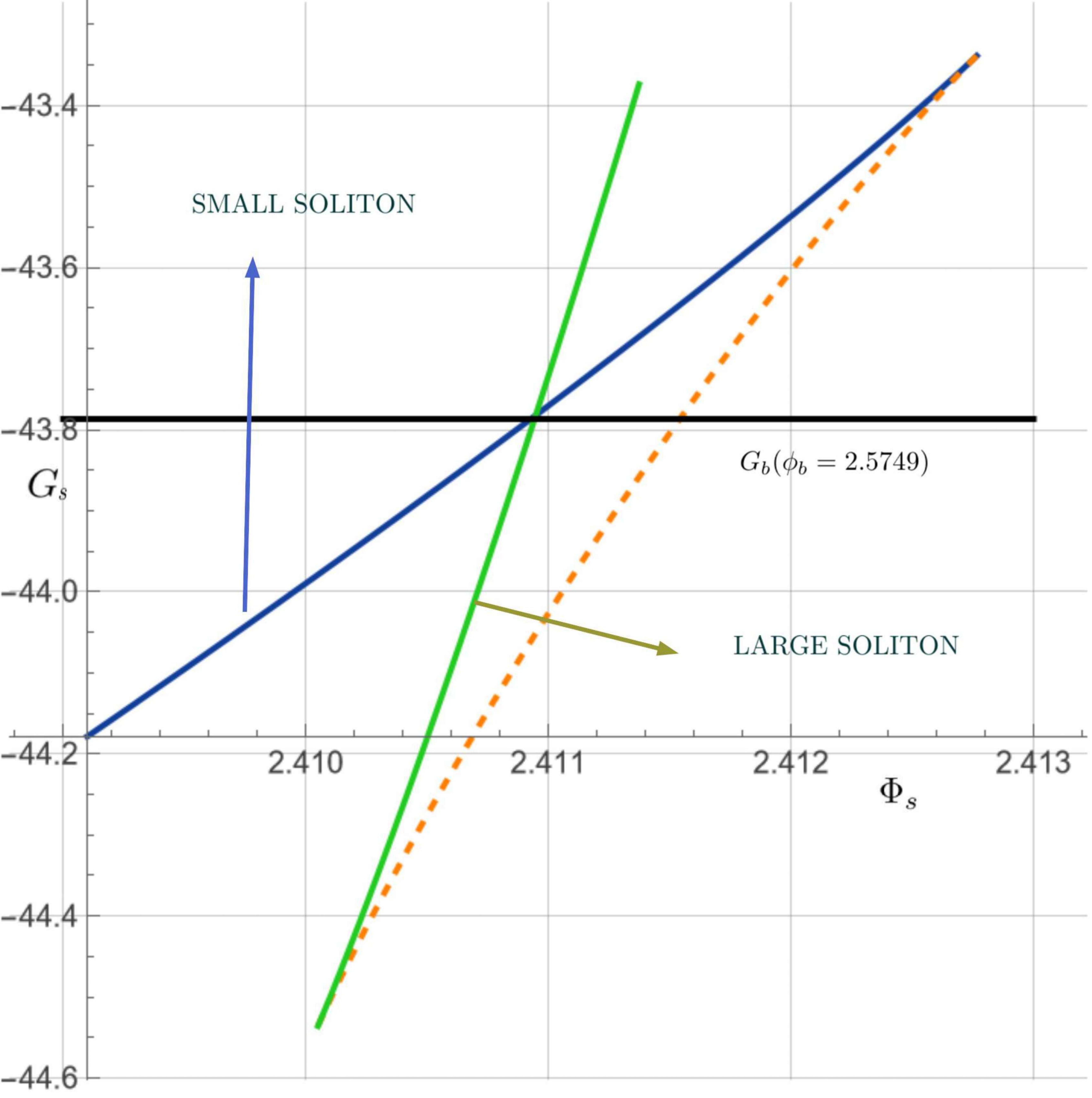}
    \caption{$G_s$ as a function of $\Phi_s$.}
\end{subfigure}

\vspace{-0.2cm}

\caption{Here, we obtain another similar configuration for $\beta_1 = 1.086879913$ and $\beta_2 = -0.7303433778$, where the small and large solitons are represented by the blue and green curves, respectively. In this case, $T_b=1.17344839$ and $\phi_b=2.57494797$. As before, $\eta=\kappa=L=0.5=\ell/2$.}
\label{fig2}
\end{figure*}

By contrast, multicritical points, characterized by the condition $G_s=G_b$, arise only for particular choices of the couplings and therefore require an additional tuning of the parameters. Physically, the extrema give rise to distinct branches of soliton solutions, each of which may correspond to a locally stable phase. Local stability is determined by the convexity properties of the Gibbs free energy, whereas global stability follows from selecting the configuration with the lowest Gibbs free energy.

\subsection{Triple Points}

As a first example, we reproduce the results of \cite{Quijada:2023fkc}. In our notation, the couplings appearing in Eq.~(\ref{eq:case1}) are related to those of \cite{Quijada:2023fkc} through $\beta_1=\alpha_1$ and $\beta_2=-\alpha_2$. By the locations of the local extrema of $\Phi_s$ and $G_s$ as functions of $r_0$ (see Figs.~\ref{fig1}(a) and \ref{fig1}(b)), it is possible to generate a swallowtail structure in the soliton Gibbs free energy, as shown in Fig.~\ref{fig1}(c). In this case, the solitonic configuration exhibits a PT from a small (blue) to a large (green) branch, where the dotted segment represents a region with negative generalized specific heat. The local extrema of $\Phi_s$ and $G_s$ determine the turning points of the swallowtail and separate branches corresponding to distinct thermodynamic phases. As a consequence, multiple locally stable soliton solutions may coexist for the same value of the magnetic flux. Additionally, when considering the BH phase, for a specific value of the electric potential $\phi_b$, the Gibbs free energies $G_b$ and $G_s$ coincide, indicating the coexistence of three configurations: one BH and two locally stable solitons. In this situation, the soliton free energy $G_s$ is insensitive to variations of the electric potential $\phi_b$, whereas the black hole free energy $G_b$ is insensitive to variations of the magnetic flux $\Phi_s$. \\

For the sake of completeness, the matter source (\ref{eq:NLE}) defined by the polynomial function (\ref{structural-function}) also admits additional analogous configurations. One such example is presented in Fig.~\ref{fig2}. As in the previous case, the magnetic flux and the soliton free energy exhibit a local maximum and a local minimum as functions of $r_0$ (Figs.~\ref{fig2}(a) and \ref{fig2}(b)), giving rise to a swallowtail structure in the free energy and the coexistence of two locally stable soliton branches (Fig.~\ref{fig2}(c)). These branches correspond to the small (blue) and large (green) solitons, respectively. Furthermore, for a particular value of the electric potential, the Gibbs free energy of the BH coincides with that of the two soliton branches, giving rise to a triple point where the three configurations coexist.

\begin{figure*}[t]
\centering

\begin{subfigure}[c]{0.30\textwidth}
    \centering
    \includegraphics[width=\linewidth]{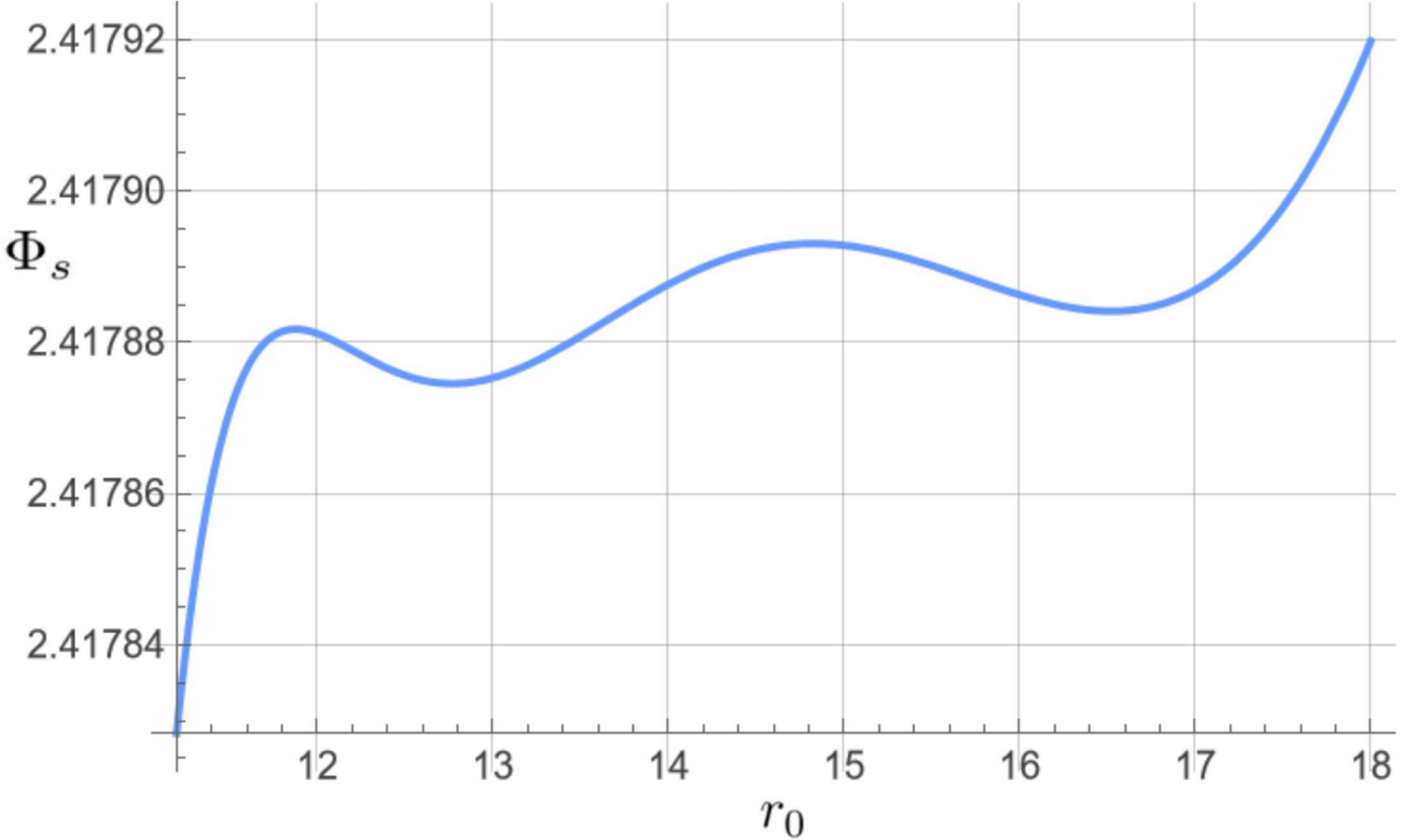}
    \caption{$\Phi_s$ as a function of $r_0$.}
\end{subfigure}
\hfill
\begin{subfigure}[c]{0.30\textwidth}
    \centering
    \includegraphics[width=\linewidth]{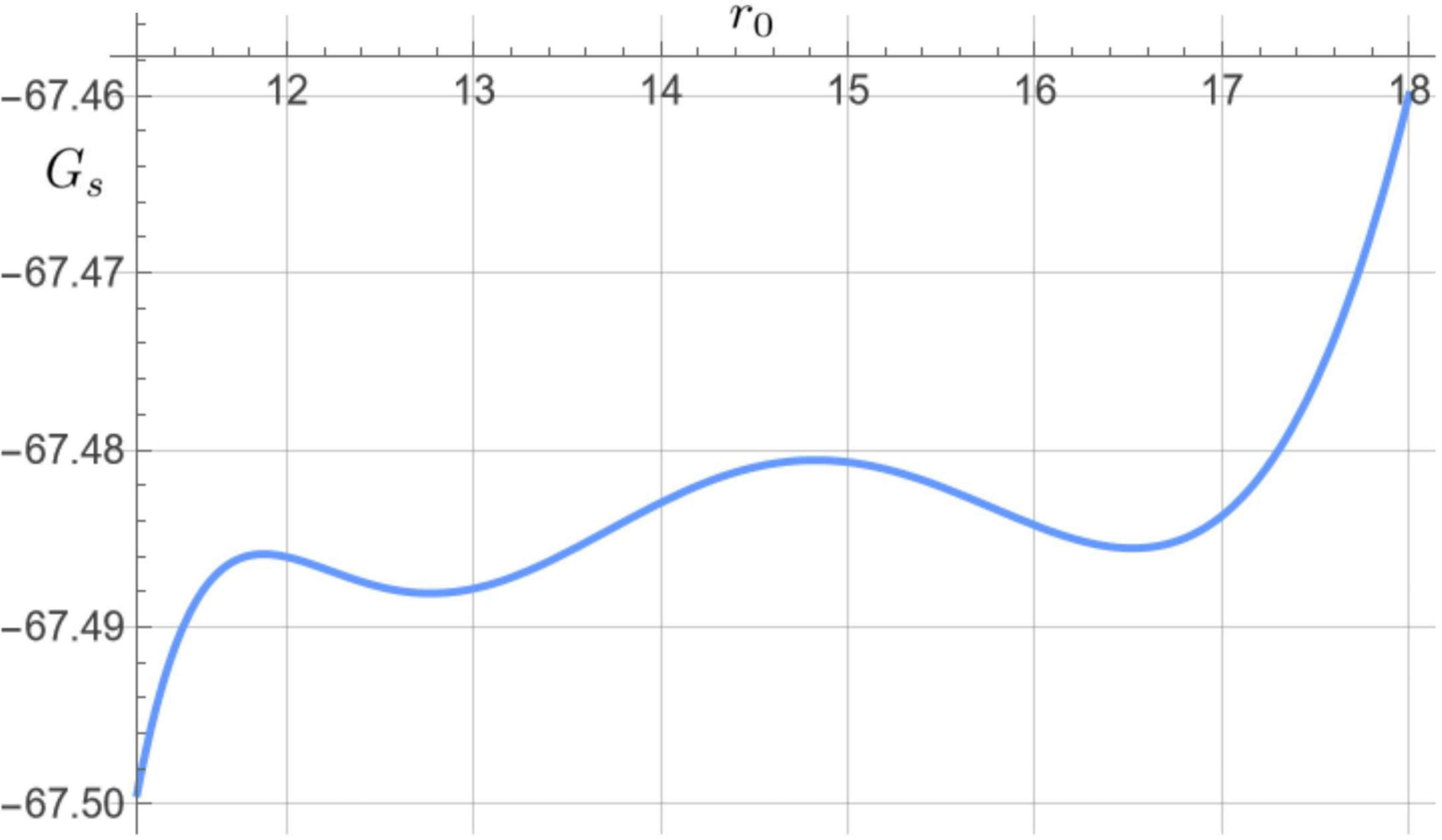}
    \caption{$G_s$ as a function of $r_0$.}
\end{subfigure}
\hfill
\begin{subfigure}[c]{0.30\textwidth}
    \centering
    \includegraphics[width=\linewidth]{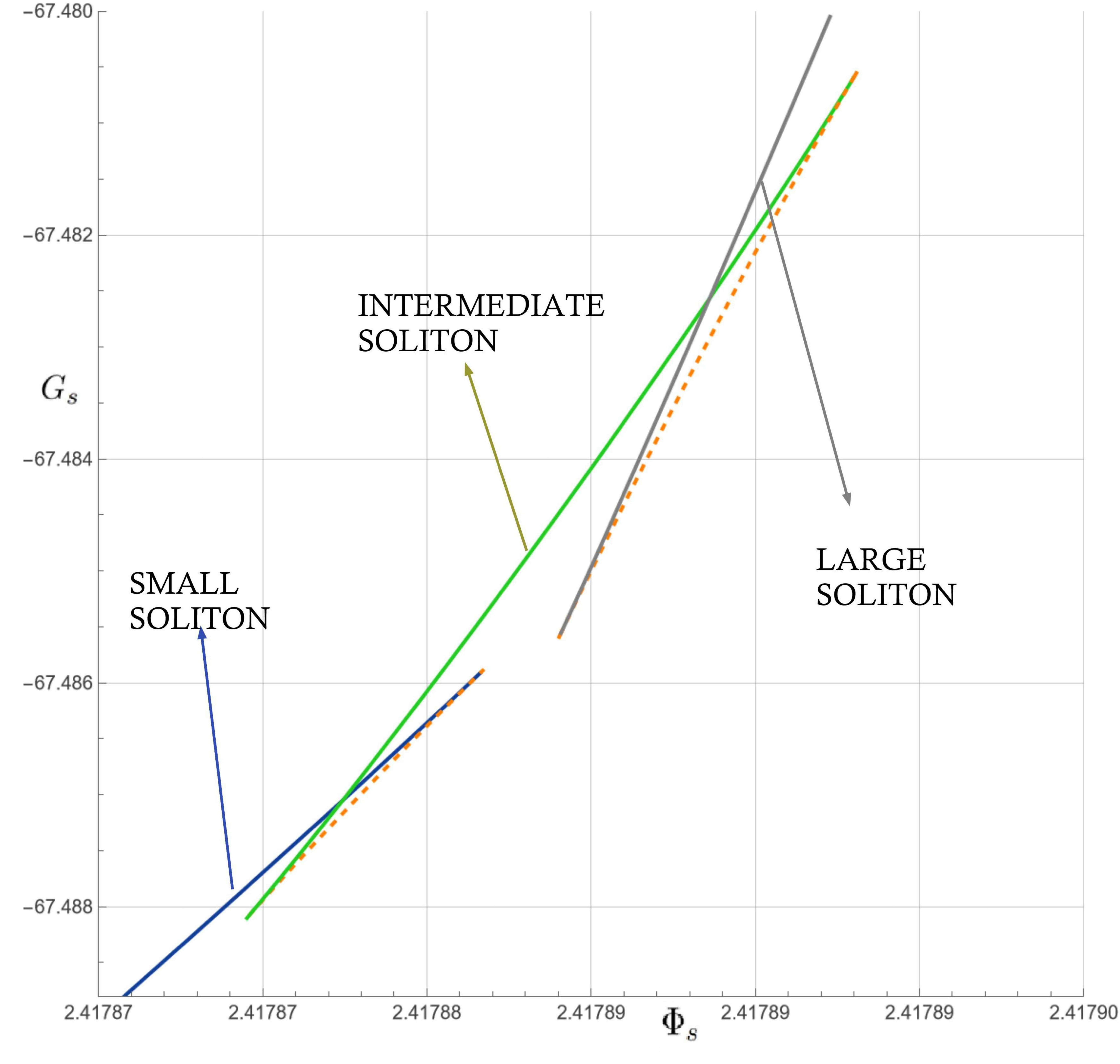}
    \caption{$G_s$ depending on $\Phi_s$.}
\end{subfigure}

\vspace{-0.2cm}

\caption{For $\beta_1 = 1.487110929$, $\beta_2 = -2.350007006$, $\beta_3 = 0.8034236$, and $\beta_4 = -0.109467$, it is possible to obtain small, intermediate, and large solitons, represented by the blue, green, and gray curves, respectively. For this configuration, $\eta=\kappa=L=0.5=\ell/2$.}
\label{fig31}
\end{figure*}
\begin{figure*}[t]
\centering

\begin{subfigure}[c]{0.30\textwidth}
    \centering
    \includegraphics[width=\linewidth]{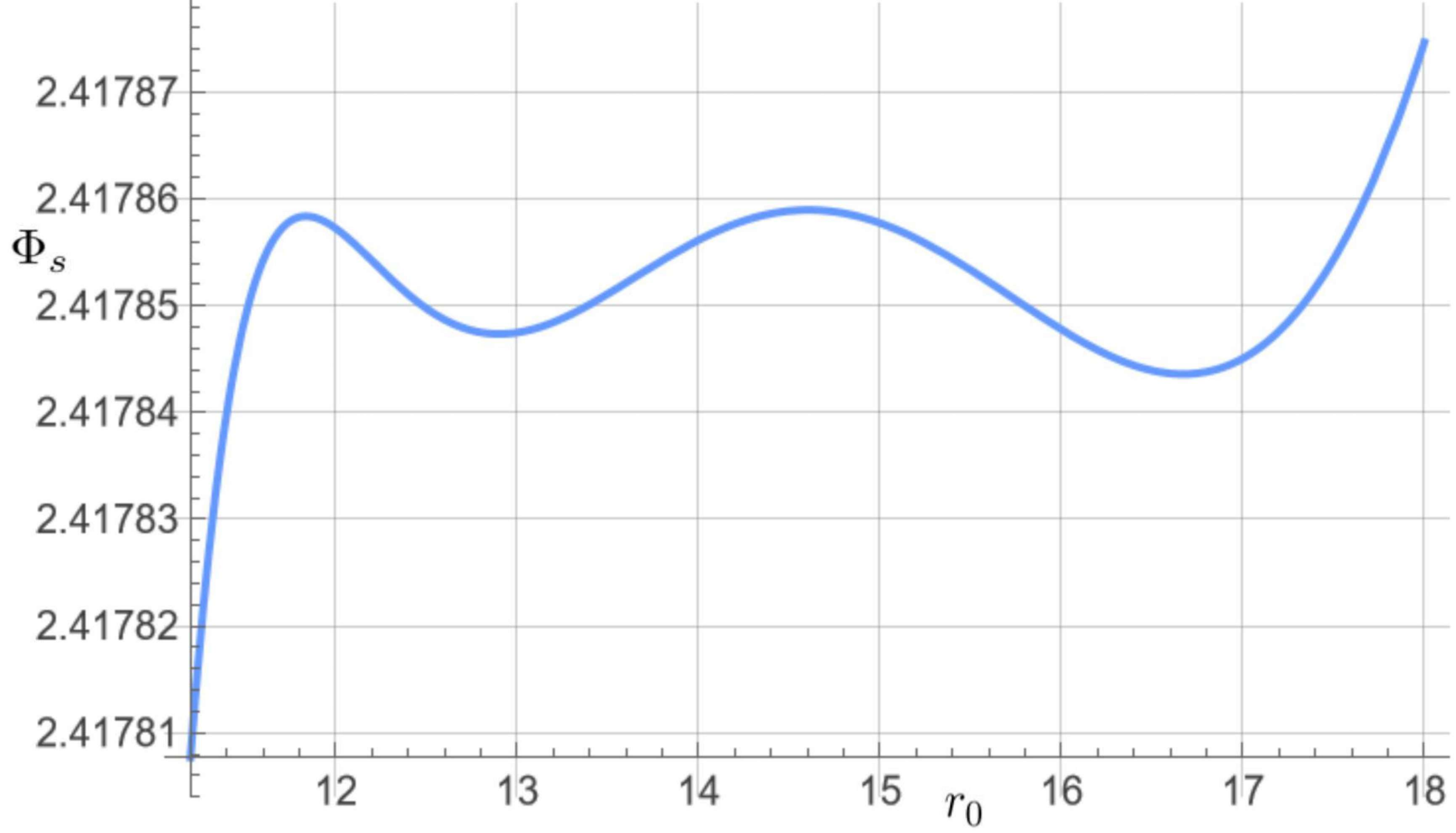}
    \caption{$\Phi_s$ as a function of $r_0$.}
\end{subfigure}
\hfill
\begin{subfigure}[c]{0.30\textwidth}
    \centering
    \includegraphics[width=\linewidth]{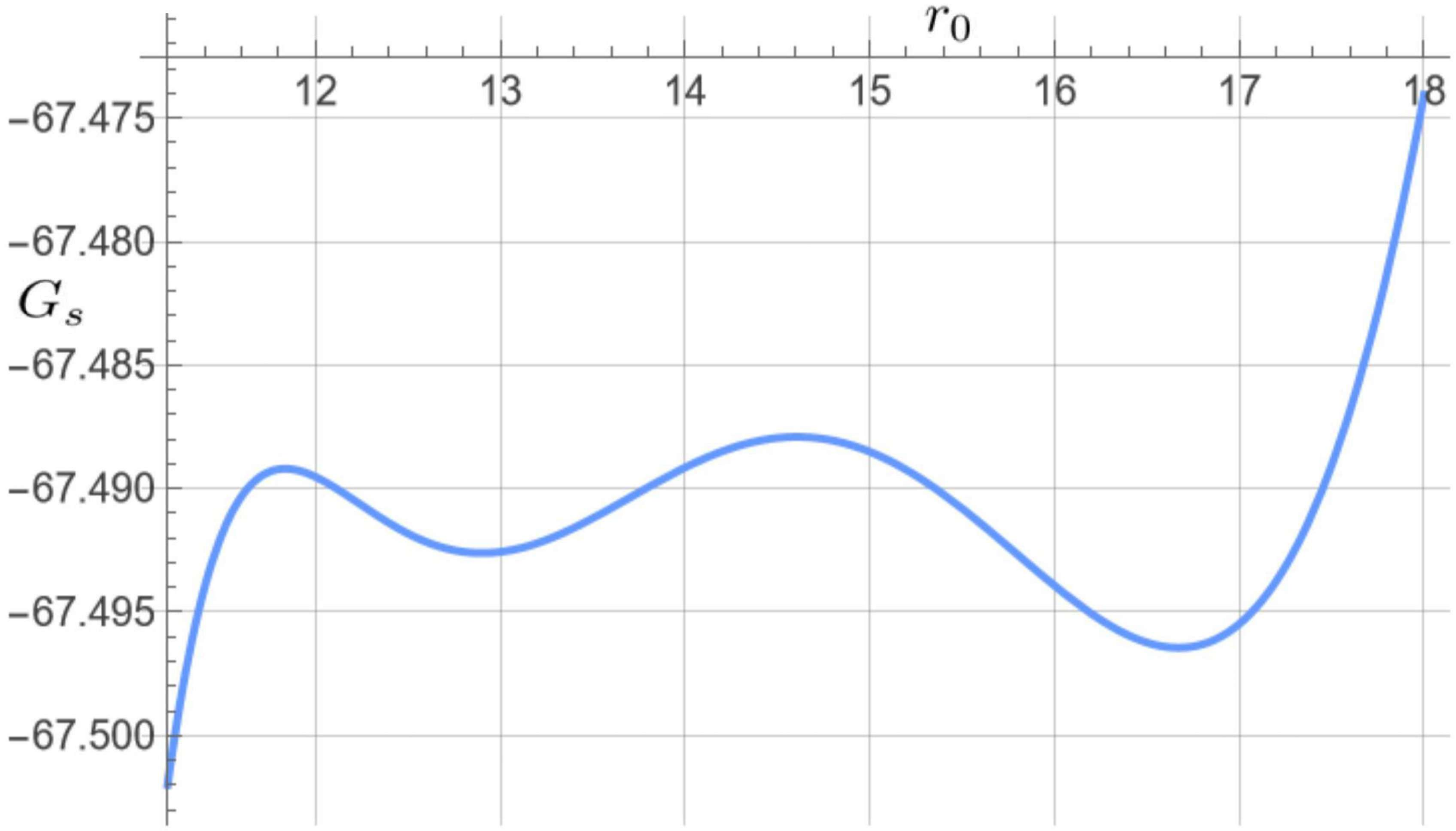}
    \caption{$G_s$ as a function of $r_0$.}
\end{subfigure}
\hfill
\begin{subfigure}[c]{0.30\textwidth}
    \centering
    \includegraphics[width=\linewidth]{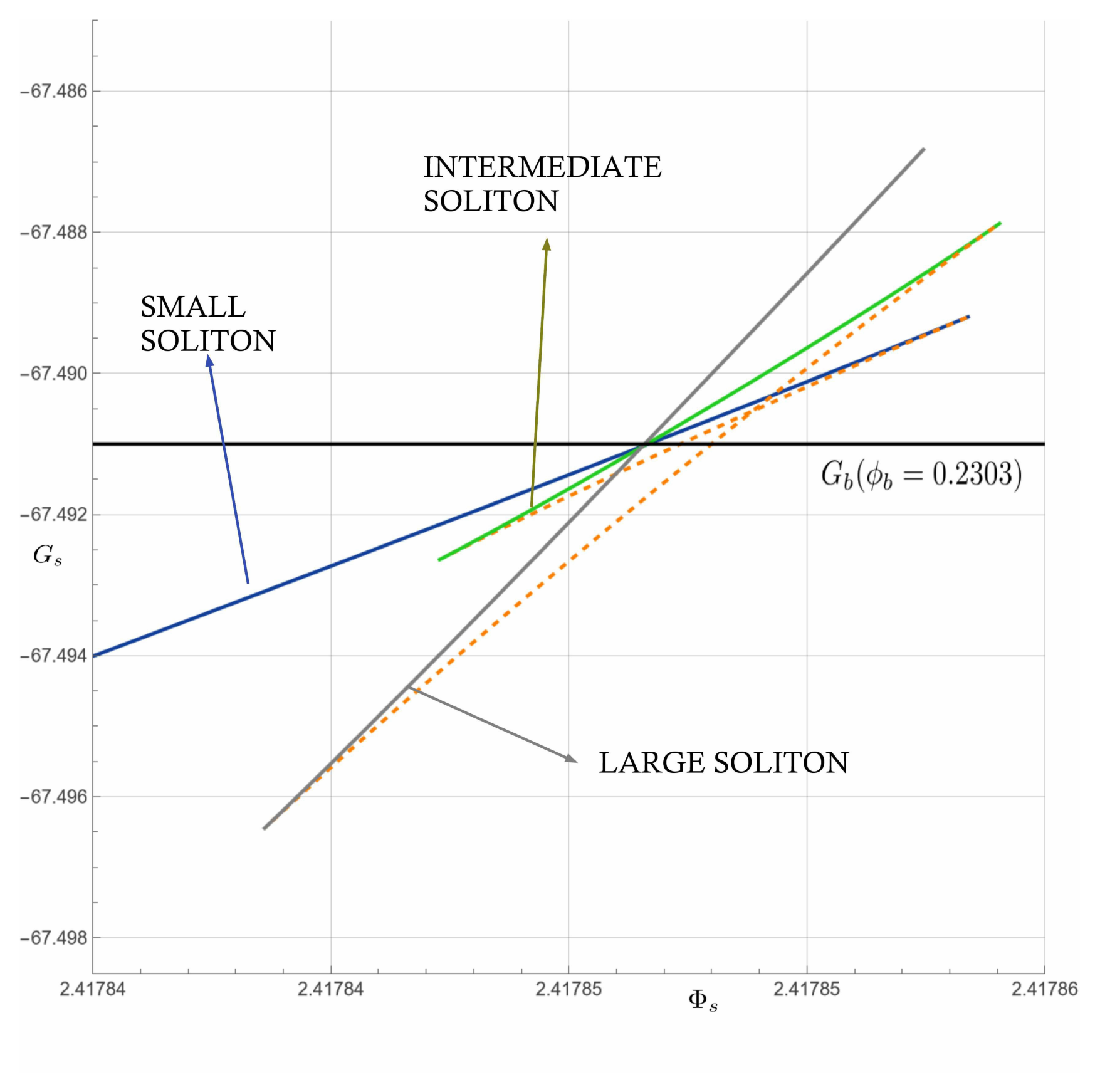}
    \caption{$G_s$ depending on $\Phi_s$.}
\end{subfigure}

\vspace{-0.2cm}

\caption{For the parameter set $\beta_1=1.487111$, $\beta_2=-2.3500178$, $\beta_3=0.803424$, and $\beta_4=-0.1094667$, another coexistence phase emerges, consisting of one BH and three locally stable solitons. The small, intermediate, and large soliton branches are represented by the blue, green, and gray curves, respectively. In this configuration, $\eta=\kappa=L=0.5=\ell/2$, with $T_b=1.542053314$ and $\phi_b=0.2303279491$.}
\label{fig3}
\end{figure*}
\begin{figure*}[t]
\centering

\begin{subfigure}[c]{0.30\textwidth}
    \centering
    \includegraphics[width=\linewidth]{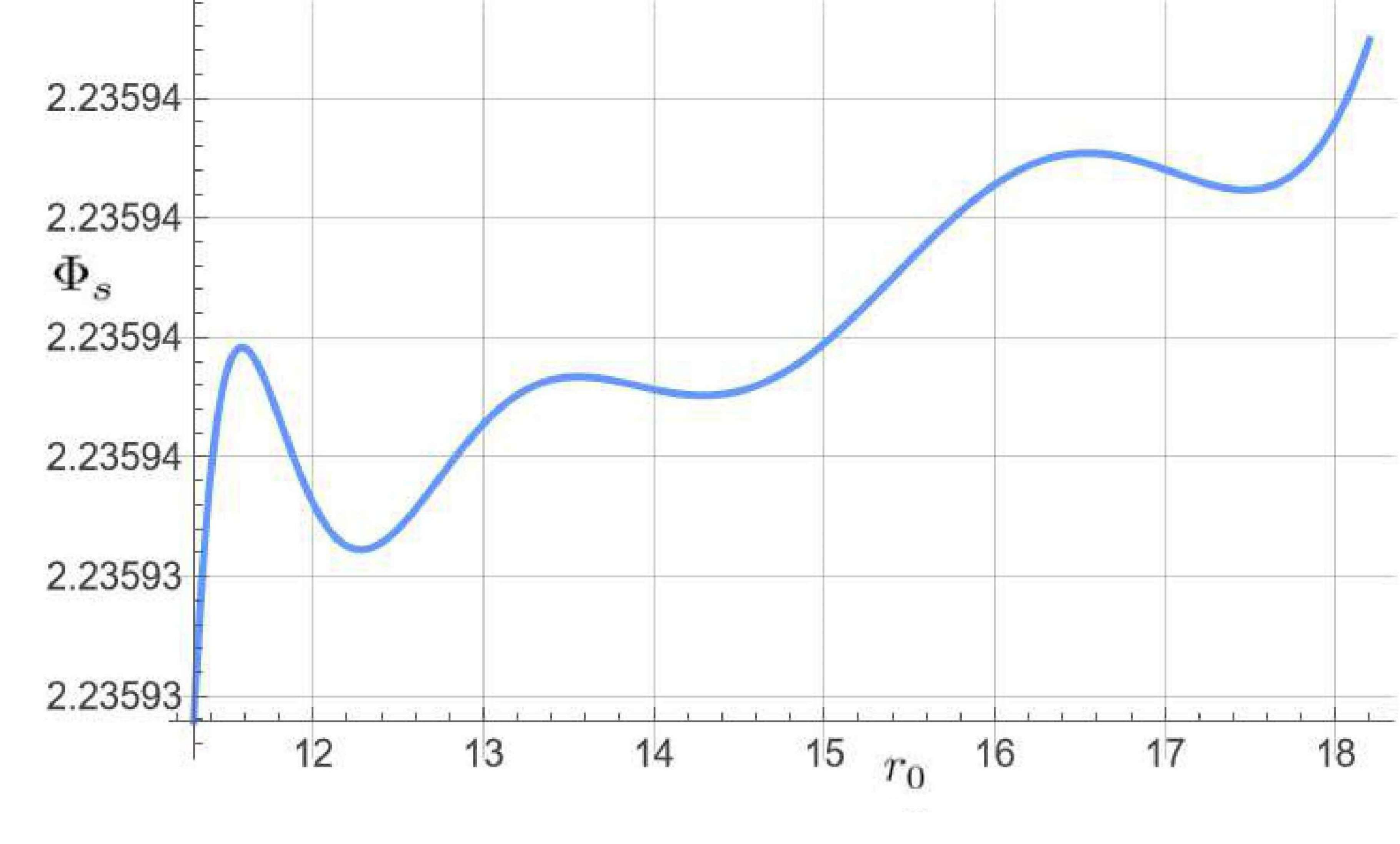}
    \caption{$\Phi_s$ as a function of $r_0$.}
\end{subfigure}
\hfill
\begin{subfigure}[c]{0.30\textwidth}
    \centering
    \includegraphics[width=\linewidth]{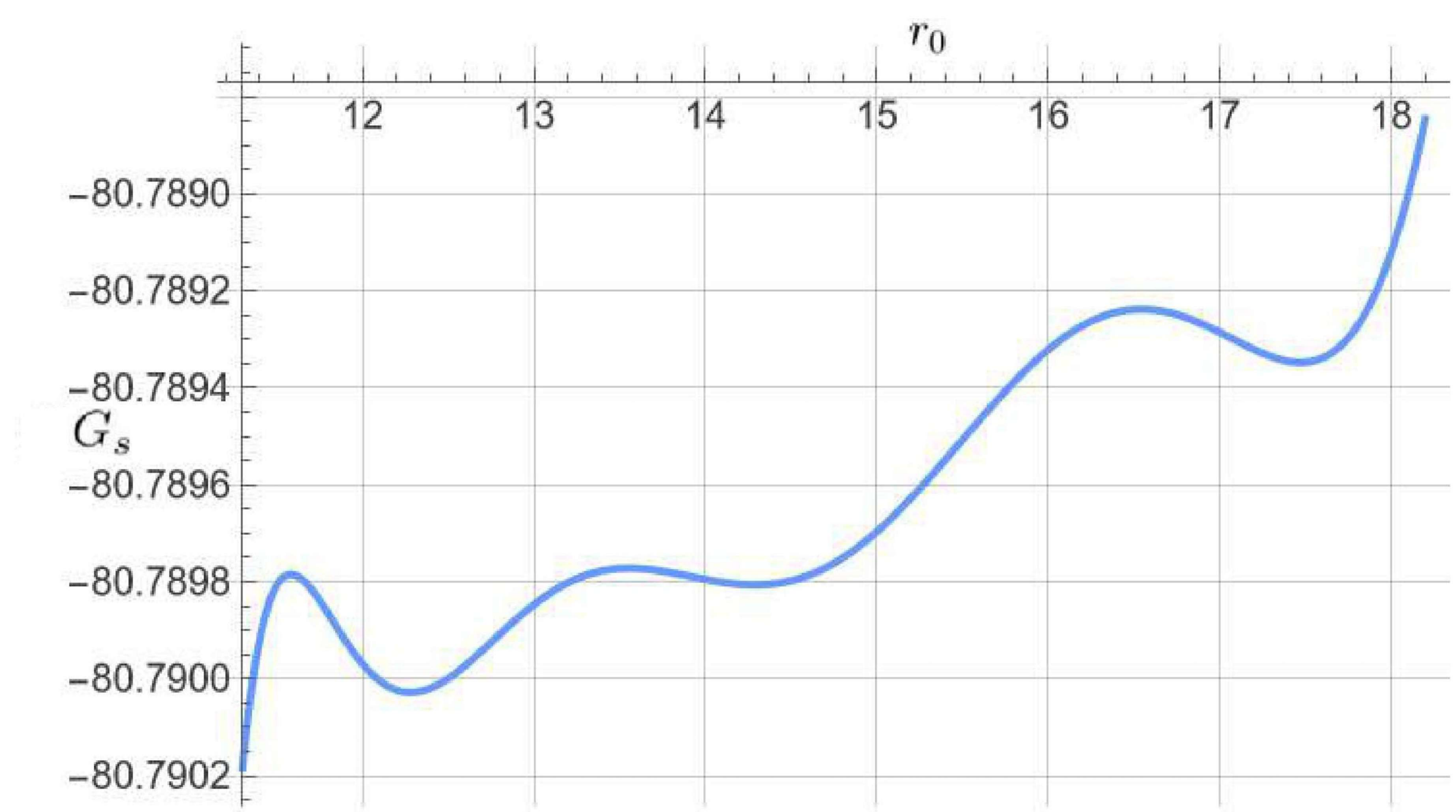}
    \caption{$G_s$ as a function of $r_0$.}
\end{subfigure}
\hfill
\begin{subfigure}[c]{0.30\textwidth}
    \centering
    \includegraphics[width=\linewidth]{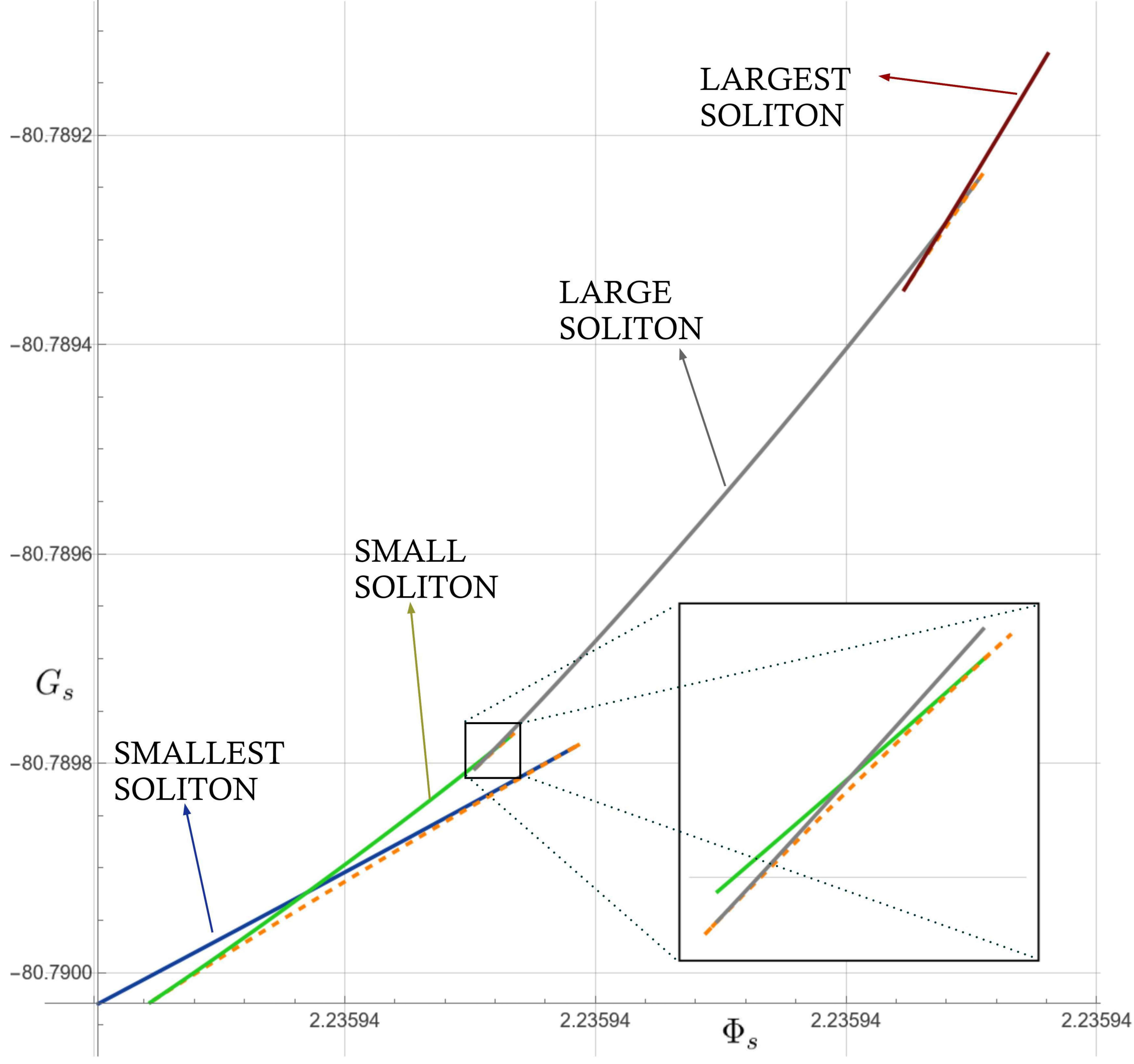}
    \caption{$G_s$ depending on $\Phi_s$.}
\end{subfigure}

\vspace{-0.2cm}

\caption{For $\beta_1 = 1.55100445$, $\beta_2 = -3.39338871$, $\beta_3 = 2.09425408$, $\beta_4 = -0.71312761$, $\beta_5 = 0.1280145940$, and $\beta_6 = -0.0097426618$, a coexistence phase emerges characterized by four locally stable solitons. The smallest, small, large, and largest soliton branches are represented by the blue, green, gray, and red curves, respectively. As before, $\eta=\kappa=L=0.5=\ell/2$.}
\label{fig41}
\end{figure*}

\begin{figure*}[t]
\centering

\begin{subfigure}[c]{0.30\textwidth}
    \centering
    \includegraphics[width=\linewidth]{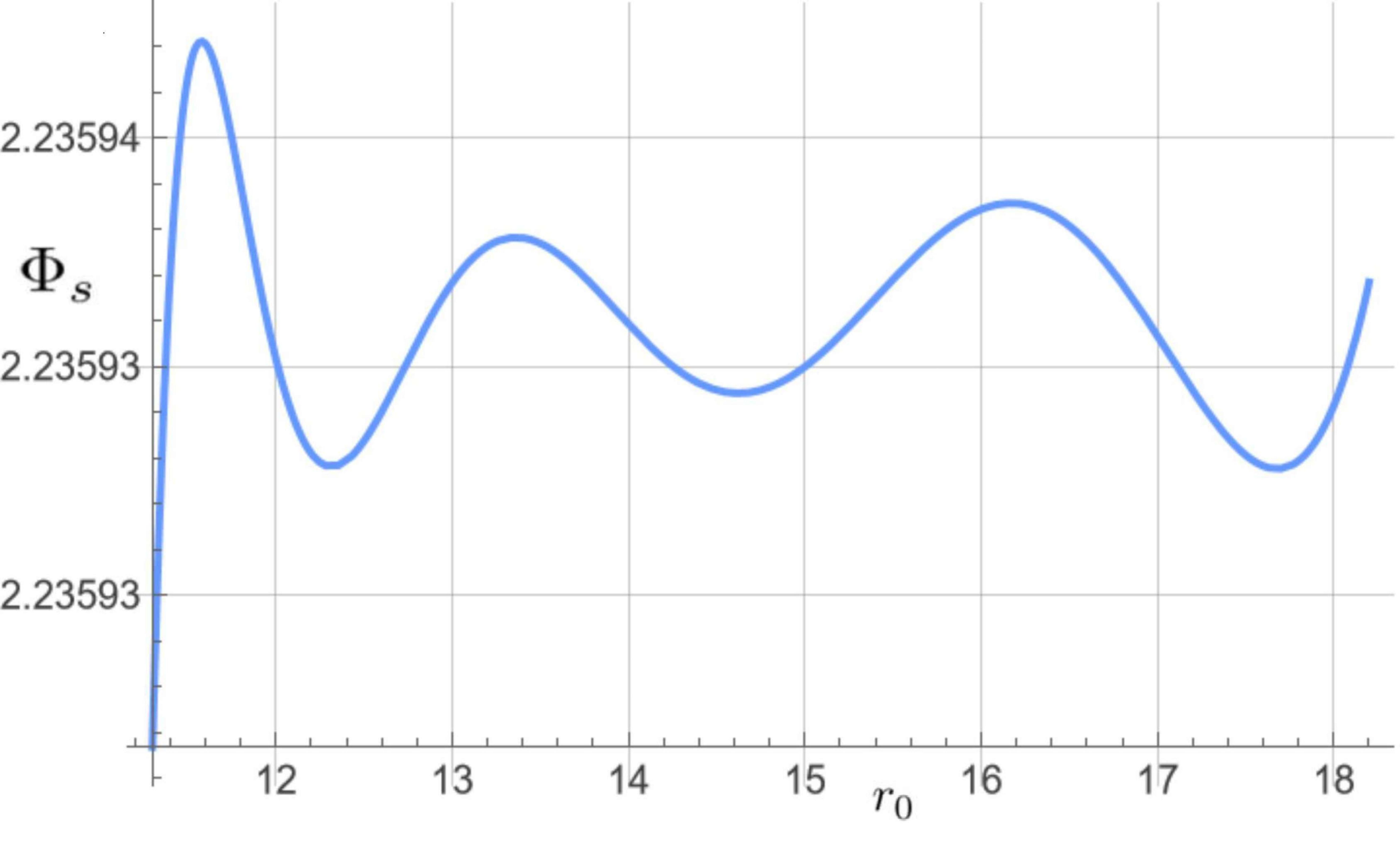}
    \caption{$\Phi_s$ as a function of $r_0$.}
\end{subfigure}
\hfill
\begin{subfigure}[c]{0.30\textwidth}
    \centering
    \includegraphics[width=\linewidth]{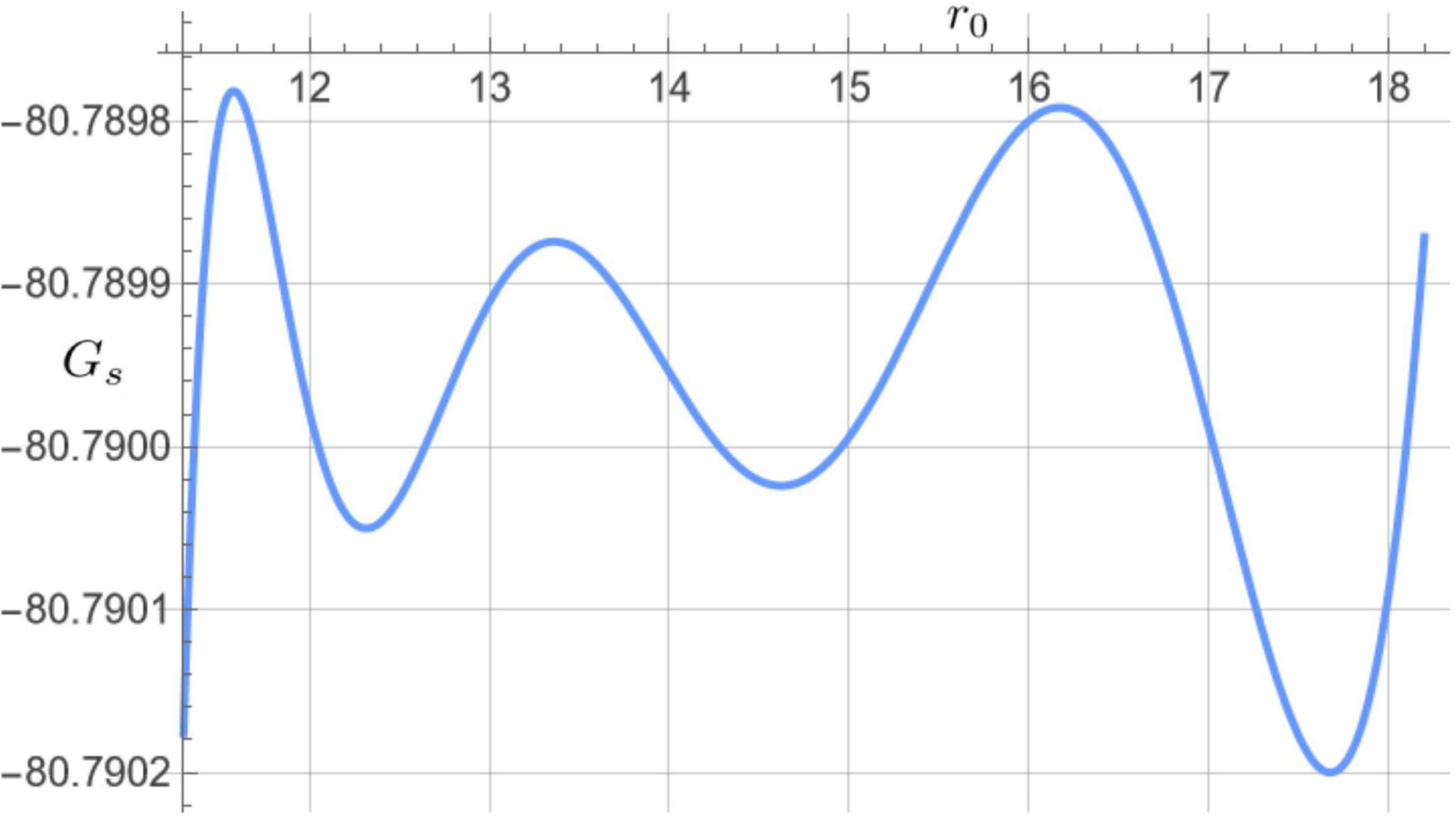}
    \caption{$G_s$ as a function of $r_0$.}
\end{subfigure}
\hfill
\begin{subfigure}[c]{0.30\textwidth}
    \centering
    \includegraphics[width=\linewidth]{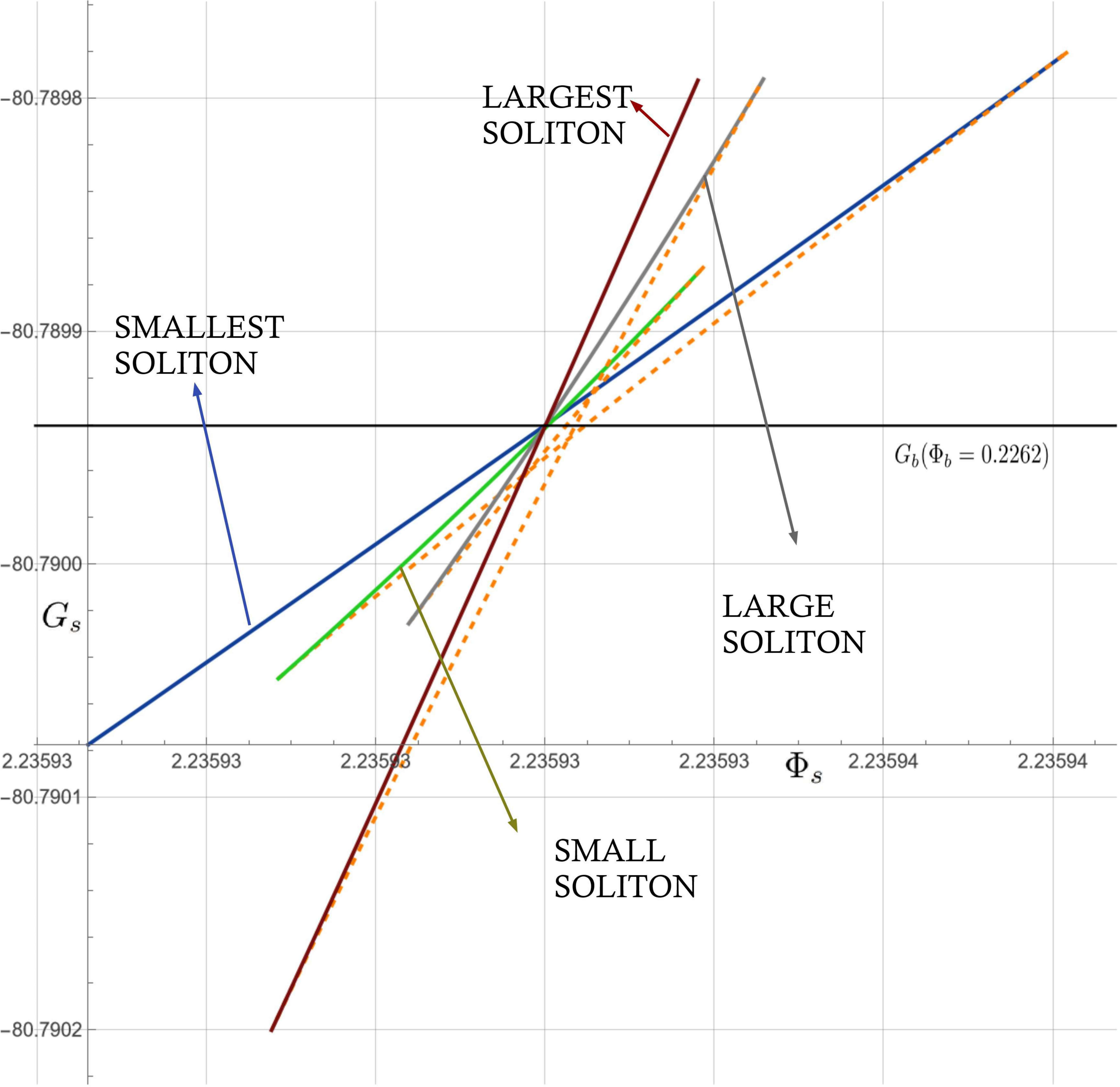}
    \caption{$G_s$ depending on $\Phi_s$.}
\end{subfigure}

\caption{For the parameter set $\beta_1 = 1.551004145$, $\beta_2 = -3.393388119$, $\beta_3 = 2.094254127$, $\beta_4 = -0.71312761255$, $\beta_5 = 0.1280145595$, and $\beta_6 = -0.0097426666$, another coexistence phase emerges, characterized by one BH and four locally stable solitons. The smallest, small, large, and largest soliton branches are represented by the blue, green, gray, and red curves, respectively. In this configuration, $\eta=\kappa=L=0.5=\ell/2$, with $T_b=1.6374984803$ and $\phi_b=0.2262481332$.}
\label{fig4}
\end{figure*}

\subsection{Quadruple Points}

From the previous analysis, it follows that increasing the degree $n$ of the polynomial function (\ref{structural-function}) allows for the generation of additional extrema. The key observation is that higher-degree polynomials introduce an additional extrema in the functions $\Phi_s$ and $G_s$ (subject to eq.~(\ref{eq:g-condition})), giving rise to multiple swallowtail structures in the free energy. \\

For instance, for $n=4$, two distinct swallowtails can be obtained, as shown in Fig.~\ref{fig31}. In this case, the solitonic sector exhibits three locally stable branches, represented by the blue, green, and gray curves, corresponding to the small, intermediate, and large solitons, respectively. Furthermore, by continuously adjusting the locations of the extrema through the couplings $\beta_i$, the two swallowtails approach each other. For the parameter configuration shown in Fig.~\ref{fig3}, they merge at a single point where the free energies of the three stable solitonic branches and the black hole coincide, giving rise to a quadruple point. \\

At this point, four distinct phases coexist in thermodynamic equilibrium, extending the triple-point structure discussed previously and demonstrating how additional extrema in the soliton free energy and magnetic flux give rise to increasingly complex coexistence phenomena.

\subsection{Quintuple points}
This pattern continues for higher values of $n$. For instance, for $n=6$, three distinct swallowtail structures can be obtained in the soliton free energy, as shown in Fig.~\ref{fig41}. In this case, the solitonic sector exhibits four locally stable branches, represented by the blue, green, gray, and red curves, corresponding to the smallest, small, large, and largest solitons, respectively. The presence of three swallowtails reflects the existence of three first-order phase transitions connecting these four stable soliton phases. Furthermore, by continuously adjusting the locations of the extrema through the couplings $\beta_i$, the swallowtails approach each other and eventually merge. For the parameter configuration shown in Fig.~\ref{fig4}, the free energies of the four stable solitonic branches and the black hole coincide, giving rise to a quintuple point.\\

In general, the number of locally stable solitons depends on the polynomial degree $n$. For even values of $n$, the Gibbs free energy $G_s$ can display up to $n/2$ swallowtail structures. By tuning the parameters so that $G_b$ overlaps with $G_s$, a $(n+4)/2$ -point configuration can be achieved, allowing the existence of $(n+2)/2$ solitons and a BH. \\

We can also inquire about the possible emergence of swallowtail structures in the Gibbs free energy $G_b$ (\ref{eq:Gb}), when analyzed as a function of   the electric potential $\phi_b$ (\ref{eq:phib}), by constraining:
\begin{equation}\label{eq:gb}
g_b(r_+,Q)=\left(\frac{3 r_+}{4 \pi \ell^2} + \frac{1}{8 \pi} \sum_{i=1}^{n} \frac{ \beta_i (-2Q^2)^i}{r_+^{4i-1}}\right)-T_0=0,
\end{equation} 
 with $T_0 > 0$, analogous to the soliton construction, where extrema of $\Phi_s$ and $G_s$ are found subject to the regularity constraint $g(r_0,q)=0$ (see eq. (\ref{eq:g-condition})). Here, the corresponding problem is to find the extrema of {$G_b=M_b-T_b S_b-\phi_b Q_b$ and $\phi_b$, subject to (\ref{eq:gb}). Here, 
 \begin{eqnarray*}
  M_b&=& \frac{2L \eta_b M}{\kappa \ell^2}, \\
  S_b&=&\frac{2L \eta_b \pi r_+^2}{\kappa \ell^2}, \\
  Q_b&=&\frac{2L \eta_b Q}{\kappa \ell^2},
 \end{eqnarray*}
are the mass, entropy, and electric charge, respectively.}
 
 {The above can be analyzed, requiring the
existence of $r_+^{\text{ext}}>0$ satisfying
\begin{eqnarray}\label{eq:sw-cond}
\frac{d \phi_b}{ d r_+} \Bigg{|}_{r_+^{\mbox{\tiny{ext}}}}=0=\frac{d G_b}{d r_+} \Bigg{|}_{r_+^{\mbox{\tiny{ext}}}},\\
\left(\frac{d^2 \phi_b}{d r_+^2} \Bigg{|}_{r_+^{\mbox{\tiny{ext}}}}\right)\left(\frac{d^2 G_b}{d r_+^2} \Bigg{|}_{r_+^{\mbox{\tiny{ext}}}}\right)>0,\nonumber
\end{eqnarray}
where, as before, all derivatives are total derivatives along the curve defined by \eqref{eq:gb}, with $Q=Q(r_+)$. Differentiating along the curve $g_b=0$ and considering the first law $\delta M_b=T \delta S_b+\phi_b \delta Q_b$ 
\begin{equation}\label{eq:key}
  \frac{dG_b}{dr_+}
  \;=\;
  -Q_b\left(\frac{d\phi_b}{dr_+}\right).
\end{equation}
Nevertheless, differentiating \eqref{eq:key} once more along the curve and evaluating at $r_+^{\text{ext}}$, one obtains
\begin{equation*}
  \frac{d^2 G_b}{dr_+^2}\bigg|_{r_+^{\text{ext}}}
  \;=\;
  -Q_b\left(\frac{d^2\phi_b}{dr_+^2}\right)\bigg|_{r_+^{\text{ext}}},
\end{equation*}
and for $Q_b>0$
\begin{equation}\label{eq:second}
  \left(\frac{d^2 \phi_b}{d r_+^2} \Bigg{|}_{r_+^{\mbox{\tiny{ext}}}}\right)\left(\frac{d^2 G_b}{d r_+^2} \Bigg{|}_{r_+^{\mbox{\tiny{ext}}}}\right)=
  -Q_b\left(\frac{d^2\phi_b}{dr_+^2}\right)^2\bigg|_{r_+^{\text{ext}}}<0.
\end{equation}
It holds for any degree $n$ of the
polynomial structural function \eqref{structural-function}, for any values of the
couplings $\beta_i$'s, and independently of the number of horizons. Consequently, no swallowtail structure can develop in $G_b(\phi_b)$ for this case.} \\
 
This result highlights a fundamental asymmetry between the two sectors. 
{The regularity constraint \eqref{eq:g-condition} implies that $\eta_s$ is a function of the soliton radius $r_0$, yielding
 a rich structure with multiple extrema and first-order transitions. However the corresponding parameter $\eta_b$ in the  BH branch  is independent of $r_+$, and so 
 under the condition (\ref{eq:gb}) cannot produce analogous extrema.} This asymmetry can be traced back to the different roles played by the electric and magnetic sectors in the Pleba\'nski formulation. The nonlinear magnetic contributions introduce additional structure in the soliton sector, whereas the electric BH sector does not develop the same level of complexity. \\

In summary, the phase structure of the system is governed by the interplay between the NLE and the soliton geometry. The polynomial degree $n$ acts as a control parameter that determines the number of coexisting phases, allowing for the systematic construction of multicritical configurations in four-dimensional gravity. This provides a novel realization of multicritical behavior in gravitational systems, where multiple phases can coexist in equilibrium.

\section{Conclusions and discussions }\label{conclusions}

In this work, we present the first concrete realization of multicritical points in four-dimensional general relativity using Pleba\'nski's nonlinear electrodynamics given in eqs. (\ref{eq:NLE})-(\ref{structural-function}). Our construction provides a systematic mechanism for  generating  multicritical behavior, where the degree of the polynomial structural function controls the number of coexisting phases.

Firstly, we established a mapping between two formulations of NLEs, the first one expressed as a power series in the Maxwell invariant $F$ (see eq. (\ref{eq:NLE-Gao})) and the other in terms of the Pleba\'nski polynomial structural function $ \mathcal{H}(P)$ given in eq. (\ref{structural-function}). This mapping shows us how the sets of couplings $\alpha_i$ and $\beta_j$ are related, recovering the cases studied previously in \cite{Gao:2021kvr}.

We used the structural function (\ref{structural-function}) to construct new families of electrically charged AdS BHs with planar horizons and magnetically charged AdS solitons. We then derived their thermodynamic quantities in the grand canonical ensemble, in order to study the corresponding Gibbs free energies $G_b$ and $G_s$. Our analysis revealed several coexistence phenomena between these phases. Notably, we observed swallowtail structures in $G_s$, which indicate first-order phase transitions within the solitonic sector. We also identified parameter values where $G_s$ and $G_b$ intersect, indicating coexistence between a BH and multiple locally stable solitons. Here, the polynomial degree $n$ takes a central role, allowing us to relate the number of independent couplings, the horizon structure, and even the existence of multicritical points.  

A key result of our study is that the number of coexisting phases is determined by the polynomial degree $n$, leading to a clear organizing principle for the emergence of multicriticality: the solitonic free energy $G_s$ can exhibit up to $n/{2}$ swallow-tail structures. By further adjusting the parameters to align $G_s$ with $G_b$, one can achieve a configuration with $(n+4)/{2}$ points, which corresponds to the coexistence of $(n+2)/{2}$ solitons alongside a single BH. We showed this mechanism explicitly for $n = 2, 4,$ and $6$ with specific choices of couplings, thereby extending previously identified triple points   \cite{Quijada:2023fkc}  to higher levels of multicriticality.

Physically, this matches the planar setting where the AdS soliton provides the confining phase; the multiple soliton branches then encode distinct confinement states competing with a deconfinement BH phase.  In contrast, studying $G_b$ as a function of the electric potential $\phi_b$  subject to (\ref{eq:gb}) shows that the necessary extremal conditions force eq. (\ref{eq:sw-cond}) at the common extrema, preventing swallowtail structures. Thus, the BH branch is insensitive to generating swallowtails, whereas the soliton sector is not. This asymmetry indicates that the NLE primarily affects the confined (solitonic) sector, while the deconfined (BH) phase remains structurally simpler.

The coexistence of multiple magnetically charged solitons with an electrically charged BH offers a gravitational realization of a generalized Gibbs phase rule \cite{Sun:2021}:
\begin{equation}\label{eq:Gibbsphase}
F = W-P + 1,
\end{equation}
where $P$ is the number of coexistent phases, $W$ is the number of thermodynamic conjugate pairs, and $F$ is the number of independent intensive parameters. Considering $W=C+1$, we recover the Gibbs phase rule, with $C$ given by the number of independent control parameters. For our situation, we note that $P=(n+4)/2$, which represents the existence of $(n+2)/2$ solitons and a BH, while an application of this rule requires careful counting of the truly independent degrees of freedom, which, in our case, is nontrivial due to the geometric regularity constraint on the soliton. For the BH sector, we have 
two independent thermodynamic variables, the temperature $T_b$
and the electric potential $\phi_b$. Nevertheless, for the soliton sector, the regularity condition (\ref{eq:g-condition}) imposes a nontrivial relation between the magnetic charge $q$ and $r_0$, meaning that $\Phi_s$ and $q$
are not independent thermodynamic variables. This observation has a direct implication for the counting of $W$, where the regularity condition (\ref{eq:g-condition})  reduces the magnetic sector to a single degree of freedom, so the appropriate counting gives $W=2$ (or $C=1$), corresponding to the conjugate pairs $(T_b,S_b)$ and $(\phi_b, Q_b)$. This does not mean $\Phi_s$ is irrelevant; it is, in fact, the quantity held fixed at the boundary when comparing competing phases in the grand-canonical ensemble. Rather, the point is that its value is not freely specifiable independently of the soliton geometry through eq. (\ref{eq:g-condition}). This is a difference from the BH sector, where $r_+$ and $Q$ can be varied independently.  

From eq. (\ref{eq:Gibbsphase}) with $W=2$, the condition $F \geq 0$ requires $3-P=(2-n)/2 \geq 0$, i.e., the coexistence of two solitons and a BH. However, we note that the constraint $F \geq 0$ in the Gibbs phase rule (\ref{eq:Gibbsphase}) is a condition on the dimension of the coexistence manifold in parameter space, not on the existence of isolated coexistence points.  To accommodate these higher-order cases, several strategies can be employed. These include incorporating the pressure (related to $\Lambda$) and volume, and enhancing the coupling parameters $\beta_i$ and their conjugates, as suggested in Ref. \cite{Tavakoli:2022kmo}. In this extended framework, the polynomial couplings $\beta_i$'s play a dual role: (i) defining the matter theory and (ii) acting as additional thermodynamic handles that can be used to tune the system toward a multicritical point. A systematic development of this perspective is left for future work.

\textcolor{black}{Additionally, since the competing phases studied here are an electrically charged
BH and a magnetically charged soliton, it would be natural to revisit this construction within the recently proposed NLE studied in \cite{Avetisyan:2021heg,Avetisyan:2022zza},  which incorporate both electric and magnetic potentials on equal footing. Such duality-symmetric formulations provide an alternative representation of the nonlinearities, with a different notion of polynomiality of the interactions, and may shed light on the electric/magnetic asymmetry of the phase structure found in this work, namely, that swallowtail structures develop in the (magnetic) solitonic sector but not in the (electric) BH sector. They would also offer a natural arena to explore dyonic configurations and the imposition of discrete or continuous electric-magnetic duality symmetries on the couplings, which we leave for future investigations.}

Considering that we are working with configurations that have a planar base manifold, a natural extension of this study would be to explore spherical or hyperbolic horizons, as well as rotating geometries in dimensions different from four beyond general relativity. This would allow us to test the robustness of multicriticality when the spacetime dimension, the gravity theory, and the topology change.
From the perspective of gauge/gravity correspondence, the presence of multiple solitons alongside a single BH may correspond to multi-phase states in the dual field theory. Studying the related order parameters, transport coefficients, and entanglement structures could offer a deeper understanding of these transitions at a microscopic level. In conclusion, the NLE (\ref{eq:NLE})-(\ref{structural-function}) offers a rich framework for examining intricate thermodynamic behaviors in gravitational systems, revealing a hierarchy of multicritical phenomena that encourages further investigation from both theoretical and holographic perspectives.

\appendix

\acknowledgments
\textcolor{black}{The authors would like to thank the anonymous referee for carefully reading our manuscript and giving valuable suggestions that led to an improved version of this
work.} This work was supported in part by the Natural Sciences and Engineering Research Council of Canada.
MB is supported by Proyecto Interno  UCM-IN-25202 l\'inea regular, and FONDECYT grant 1262452.


\end{document}